\documentclass[12pt,preprint]{aastex}

\newcommand{\be}{\begin{equation}}
\newcommand{\ee}{\end{equation}}
\newcommand{\bstar}{{ \beta }}

\newcommand{\energy}{{\cal E}} 
\newcommand{\uvec}{{ {\bf u} }}
\newcommand{\bvec}{{ {\bf B} }}
\newcommand{\zhat}{{ {\hat z} }} 
 
\newcommand{\rhat}{{ {\hat r} }} 
\newcommand{\phat}{{ {\hat p} }} 
\newcommand{\thetat}{{ {\hat \theta} }} 
\newcommand{\phihat}{{ {\hat \phi} }} 
\newcommand{\epp}{{ {\underline{\epsilon}}_p}} 
\newcommand{\epq}{{ {\underline{\epsilon}}_q}} 
\newcommand{\epphi}{{ {\underline{\epsilon}}_\phi}} 

\newcommand{\fmdot}{{ {\cal F}_{\rm m} }}  
\newcommand{\efficiency}{{ \eta_{\rm rad} }} 
\newcommand{\effabsorb}{{ \eta_{\rm abs} }} 
\newcommand{\sigmauv}{{ \sigma_{\rm uv} }}
\def\lta{\,\raise 0.3 ex\hbox{$ < $}\kern -0.75 em
 \lower 0.7 ex\hbox{$\sim$}\,}
\def\gta{\,\raise 0.3 ex\hbox{$ > $}\kern -0.75 em
 \lower 0.7 ex\hbox{$\sim$}\,} 

\begin{document} 

\title{\bf MAGNETICALLY CONTROLLED OUTFLOWS \\
FROM HOT JUPITERS} 

\author{Fred C. Adams$^{1,2,3}$} 

\affil{$^1$Michigan Center for Theoretical Physics \\
Physics Department, University of Michigan, Ann Arbor, MI 48109} 

\affil{$^2$Astronomy Department, University of Michigan, Ann Arbor, MI 48109} 

\affil{$^3$Kavli Institute for Theoretical Physics, 
University of California, Santa Barbara, 93106} 

\begin{abstract} 

Recent observations that indicate that some extrasolar planets
observed in transit can experience mass loss from their surfaces.
Motivated by these findings, this paper considers outflows from Hot
Jupiters in the regime where the flow is controlled by magnetic
fields. Given the mass loss rates estimated from current observations
--- and from theoretical arguments --- magnetic fields will dominate
the flow provided that field strength near the planet is greater than
$\sim1$ gauss, comparable to the surface fields of the Sun and
Jupiter. The problem can be separated into an inner regime, near the
planet, where the outflow is launched, and an outer regime where the
flow follows (primarily) stellar field lines and interacts with the
stellar wind.  This paper concentrates on the flow in the inner
regime. For a dipole planetary field with a spatially constant
background contribution, we construct a set of orthogonal coordinates
that follow the field lines and determine the corresponding
differential operators. Under the assumption of isothermal flow, we
analytically find the conditions required for escaping material to
pass smoothly through the sonic transition, and then estimate the mass
outflow rates.  These magnetically controlled outflows differ
significantly from previous spherical models: The outflow rates are
somewhat smaller, typically ${\dot M}$ $\sim 10^{9}$ g/s, and the flow
is launched primarily from the polar regions of the planet. In
addition, if the stellar wind is strong enough, the flow could be 
reversed and the planet could gain mass from the star.

\end{abstract} 


\section{Introduction} 
\label{sec:intro}

Among the hundreds of extrasolar planets discovered to date, a
substantial fraction orbit their stars with periods of 10 days or
less. These planets are thought to have formed further out in their
solar systems (e.g., Lissauer \& Stevenson 2007), and subsequently
migrated inward (e.g., Papaloizou \& Terquem 2006) where they become
stranded at small semimajor axes, perhaps due to disk truncation (Shu
et al. 1994, Lin et al. 1996) or because the disk loses so much gas
that it can no longer move planets.  After planets reach these inner
orbits, they are subjected to intense heating from their parental
stars.  This heating, which is most effective for UV photons, can
drive photo-evaporative flows from the planetary surfaces. In the most
extreme cases, the resulting mass loss could affect both the final
masses and densities of the planets. In other cases, the outflows can
observable --- even if their effect on the final mass is modest ---
and can provide important information about planetary properties. This
paper explores outflows from Hot Jupiters in the regime where magnetic
fields are strong enough to guide the flow and thereby determine the
outflow geometry.

Mass loss from Hot Jupiters has been observed in association with the
transiting planet HD209458b (Vidal-Majar et al. 2003, 2004; see also
D{\'e}sert et al. 2008, Sing et al. 2008, Lecavelier des Etangs et
al. 2008).  A recent followup observation, using the Cosmic Origins
Spectrograph on the {\it Hubble Space Telescope} implies a mass
outflow rate ${\dot M} \approx 8 \times 10^{10}$ g/s from this planet
(Linsky et al. 2010).  In addition, signatures of atmospheric
evaporation from the extrasolar planet HD189733b have recently been
reported (Lecavelier des Etangs et al. 2010).  More examples are
expected in the term future, as well as null detections (Lecavelier
des Etangs 2007). As a result, the collection of close-in extrasolar
planets provides a laboratory to study the process of mass loss from
planetary bodies.

Theoretical models of mass loss from extrasolar planets have been
considered previously.  Pioneering models of outflows from these
planetary bodies have been constructed (Lammer et al. 2003; Baraffe et
al. 2004, 2006) and indicate that substantial mass loss can take
place. However, related studies of the effects of mass loss on the
population of close-in extrasolar planets show that it is difficult to
explain the observed mass distribution (Hubbard et al. 2007). In any
case, a number of open questions remain.  The aforementioned studies
(primarily) use simple energy-limited outflow models (see also Watson
et al. 1981), in conjunction with physically motivated scaling laws,
and produce a range of outflow rates for given planetary masses and
external UV fluxes. The next generation of theoretical calculations
considered refined treatments of the chemistry, photoionization, and
recombination (Yelle 2004; Garc{\'i}a Mu{\~n}oz 2007), as well as the
effects of tidal enhancement (Murray-Clay et al. 2009, hereafter MCM). 
More recently, two-dimensional models of planetary winds have been
considered (Stone \& Proga 2009) and indicate that the mass loss rates
can be less than those in the spherical limit. In addition to
planetary outflows, alternative explanations of the observations have
been put forth, where the inferred excess material is due to a
confined exosphere (Trammell et al. 2010; hereafter TAL) or a mass
transfer stream (Lai et al. 2010).

This theoretical investigation into planetary outflows adopts a new
approach. As outlined below, the magnetic fields -- from both the star
and the planet -- are generally strong enough to guide the flow (see
also TAL).  When the outflow follows the magnetic
fields lines, the geometry of the flow pattern is set by the field
structure, but can be quite complicated. In particular, the outflows
depart significantly from spherical symmetry and existing (primarily
spherical) models are not applicable. In spite of this complication,
the outflow problem can be reduced to one dimension by constructing a
new orthogonal coordinate system where one coordinate follows the
magnetic field lines. This approach allows for the outflow properties
to be determined semi-analytically. In this context, the term
``semi-analytic'' refers to models where the equations are reduced to,
at most, ordinary differential equations.

The problem of magnetically controlled outflows from planets can be
divided into subproblems. Since the magnetic field structure of the
planet guides the flow --- for the regime considered here --- the
field is independent of the outflow and can be determined separately.
Section \ref{sec:overview} discusses the conditions required for this
approach to be valid and provides an overview of the relevant scales
in the problem. This work assumes that the planetary field has a
dipole form. Near the planetary surface, the stellar component of the
magnetic field is smaller than the planetary field and is slowly
varying (spatially); as a starting approximation, we thus assume that
the stellar field provides a constant background contrribution. For
this geometry, we construct a set of orthogonal coordinates that
follow the field lines and hence the flow (Section \ref{sec:field}).
For a given field configuration, the outflow problem can be separated
further into two regimes: [1] The launch of the outflow near the
planetary surface; this flow depends on the heating and cooling in the
vicinity of the planet, and the flow geometry is constrained by the
magnetic field configuration. [2] The propagation of the outflow, away
from the planet, as enters the regime where the magnetic field
structure and gravitational forces are dominated by the star. This
paper focuses on the launch of the wind, in Section \ref{sec:launch},
and the propagation problem is left for future work.  The
observational implications of these outflows are discussed in Section
\ref{sec:observe}. The paper concludes, in Section \ref{sec:conclude},
with a summary of results, a discussion of their implications, and
some directions for future work.

\section{Overview} 
\label{sec:overview} 

This section defines the basic scales in the problem and justifies our
approach.  After presenting an estimate for the outflow rate, we show
that the gas is well-coupled to the magnetic field and that the
magnetic pressure is larger than the ram pressure of the flow by
several orders of magnitude. We also discuss the background magnetic
field provided by the star and the effects of the stellar wind, where
both of these effects limit the sphere of influence of the planet.  
A background magnetic field component (from the star) is included,
whereas additional contributions from currents are shown to be small.

We first make an order of magnitude estimate for the mass outflow rates
from Hot Jupiters using a simple scaling argument: If we assume that
the outflow is limited by the rate at which the gas gains energy from
the stellar UV flux, then the mechanical luminosity of the outflow 
$G M_P {\dot M} / R_P$ must balance the rate of energy deposition, 
$\efficiency F_{UV} \pi R_P^2$, where the parameter $\efficiency$
includes the efficiency of energy capture and takes into account the
fact that radiation can be absorbed above the planetary surface (at
$R_P$). The resulting mass outflow rate $\dot M$ is thus given by
\be
{\dot M} = \efficiency {\pi R_P^3 F_{UV} \over G M_P} \, , 
\label{mdotestimate} 
\ee
where this expression would be exact if one could determine the
correct value of $\efficiency$.  Next we note the coincidence that a
particle of mass $\mu$ living at the surface of a Jovian planet has
potential energy $\epsilon_\mu$ given by 
\be
\epsilon_\mu = {G M_P \mu \over R_P} \approx 13.9 \, {\rm eV} \, , 
\ee 
where we have used typical values $\mu$ = $m_P$, $M_P = 1 M_J$, and
$R_P = 10^{10}$ cm to evaluate the energy. This potential energy scale
is almost the same as the ionization energy for Hydrogen atoms. As a
result, we can write the mass outflow rate in the form
\be
{\dot M} = \left( \efficiency 
{ \langle h \nu \rangle \over \epsilon_\mu} \right) 
\mu \left( {\pi R_P^2 F_{UV} \over \langle h \nu \rangle} \right) 
\approx 1.5 \times 10^{10} \, \, {\rm g} \, \, {\rm s}^{-1} \, \, 
\left( {F_{UV} \over 450 \, \, {\rm erg} \, \, {\rm s}^{-1} \, \, 
{\rm cm}^{-2} } \right) \, , 
\label{mdotbalance} 
\ee
where the first (dimensionless) term in brackets is close to unity and
where the second term in brackets represents the number of UV photons
intercepted by the outflow per unit time. The UV flux is scaled to the
benchmark value $F_{UV}$ = 450 erg s$^{-1}$ cm$^{-2}$, the flux
appropriate for the quiet Sun at a distance of $a$ = 0.05 AU (e.g.,
MCM, Woods et al. 1998).  The resulting numerical estimate for the
outflow rate agrees with previous results (e.g., Garc{\' i}a Mu{\~n}oz
2007). When the efficiency $\efficiency$ is high, the outflow rate is
thus determined by an approximate balance with one outgoing particle
per incoming UV photon.

The above estimate ignores magnetic fields, whereas this paper
considers the wind to be guided by the field. In order for the plasma
to be well-coupled to the magnetic field, the cyclotron frequency
$\omega_C$ must be larger than the collision frequency $\Gamma$. The
cyclotron frequency is given by $\omega_C$ = $B q$ / $(mc)$ and the
collision rate is given by $\Gamma = n \sigma v$.  For the parameter
space of interest, we expect the magnetic field strength near the
planet to be $B \sim 1 - 40$ G (e.g., TAL) and the
collision cross section to be $\sigma \sim 2 \times 10^{-13}$ cm$^2$
(see Shu 1992, Spitzer 1978, Surrock 1994).  Because the continuity
equation implies that ${\dot M} = 4 \pi r^2 \rho v$, we can write the
ratio of frequencies in the form
\be
{\omega_C \over \Gamma} = {q B \over c m n \sigma v} = 
{4 \pi q B r^2 \over c \sigma {\dot M}} \approx 10^4
\left( {B \over 1 {\rm G}} \right) 
\left( {\sigma \over 2 \times 10^{-13} \, {\rm cm}^2} \right)^{-1} 
\left( {{\dot M} \over 10^{10} \, \, {\rm g} \, \, 
{\rm s}^{-1}} \right)^{-1}  \, .  
\label{fnumber} 
\ee 
Although the field strength decreases with distance from the planet,
equation (\ref{fnumber}) shows that the frequency ratio scales like 
$B r^2$. Since the dipole field strength $B \propto 1/r^3$, the ratio
$\omega_C / \Gamma$ decreases as one power of the radius and hence
formally exceeds unity out to a radius $r_C \sim 10^4 R_P$, well
beyond the launching radius $r_s$ of the wind (where $r_s \approx 3
R_P$; see Section \ref{sec:launch}). This estimate for $r_C$ is much
larger than the radial scale where the stellar environment ---
including the stellar wind, magnetic field, and gravity --- dominates
that of the planet. Although the number density of the wind may
continue to decrease in this regime, the magnetic field strength will
be larger than the scaling used here and the wind will remain tied to
the field.  We thus conclude that $\omega_C \gg \Gamma$ for the regime
of parameter of interest, and that the outflow is well coupled to the
magnetic field.

Another necessary condition for the magnetic field to guide the
outflow is that the magnetic pressure must be larger than the ram
pressure of the flow. Here we find the radius where the two pressures
are equal, i.e., where $\rho v^2 \approx$ $B^2/8\pi$. We can write the
density in terms of the outflow rate ${\dot M} \approx 4 \pi r^2 \rho
v$ = $\fmdot 4 \pi r^2 \rho a_s$, where $a_s$ is the sound speed and
where the second equality defines the parameter $\fmdot$.  The
magnetic pressure and the ram pressure are equal when $B^2 r^2 = 2
{\dot M} a_s \fmdot$. After scaling the magnetic field using the usual
dipole relationship, so that $B = B_P (R_P/r)^3$, we can for solve the
radius within which the magnetic field is dominant, i.e.,
\be
{r \over R_P} = \left( B_P R_P \right)^{1/2} 
\left( 2 {\dot M} a_s \fmdot \right)^{-1/4} \approx 
26 \left({ B_P \over 10 \, \, {\rm G} } \cdot 
{ R_P \over 10^{10} \, \, {\rm cm} } \right)^{1/2}  
\left({ {\dot M} \over 10^{10} \, {\rm g/s} } \cdot 
{ a_s \over 10 \, {\rm km/s}} \right)^{-1/4} \, . 
\label{dominate} 
\ee
This equation is evaluated using the surface field $B_P = 10$ G, the
value estimated for extrasolar planets (Christensen et al. 2009; see
also TAL); for comparison, the surface field strength
for Jupiter is somewhat lower, $B_P = 4.2$ G (Stevenson 2003).  This
result implies that the magnetic pressure exceeds the ram pressure of
the outflow for radii near the planet, where ``near'' is defined to be
within about 26 $R_P$. However, the stellar magnetic field is
generally stronger than the planetary field at these radii, so that
the magnetic field pressure dominates at all distances from the
planet. In addition, the sonic transitions take place at $r \sim 3
R_P$, well within the boundary defined by equation (\ref{dominate}).
Using the dipole scaling relation for the field stength, we find that
the magnetic pressure is larger than the ram pressure of the outflow
by a factor of $\sim10^4$ at the sonic surface and by a factor of
$\sim10^6$ at the planetary surface. As a result, the outflow must be
magnetically controlled. For completeness, we note that these field
configurations contain an X-point, a location where the field vanishes
in the equatorial plane and where the condition (\ref{dominate}) for
magnetic pressure domination fails; this complication affects only a
few streamlines and reduces the overall mass outflow rate by a small
amount.

The star supports a stellar wind that also provides a ram pressure.
In order for the planet to successfully launch an outflow under
magnetically controlled conditions, the ram pressure of the stellar
wind cannot be too large.  The ratio of the ram pressure of the
stellar wind to that of the planetary outflow can be written in the form 
\be
{P_{{\rm ram}\ast} \over P_{{\rm ram}P}} = 
{ ( {\dot M} v )_\ast \over ( {\dot M} v )_P } 
\left( {r \over a} \right)^2 \, , 
\label{ramratio} 
\ee
where $r$ is the radial coordinate centered on the planet and the
stellar parameters are evaluated at the location of the planet (a
distance $a$ from the star).  If the stellar wind has an outflow rate
${\dot M}_\ast = 10^{12}$ g/s and outflow speed $v_\ast$ = 400 km/s,
comparable to the values for the Sun, then the ratio of ram pressure
in equation (\ref{ramratio}) is close to unity within a few planetary
radii.  As shown above, $B^2 \gg \rho v^2$ for the planetary wind, so
that the same is true for the stellar wind; as a result, the magnetic
pressure near the planet will be much larger than the ram pressure of
both the stellar wind and the planetary wind, and the outflow will be
magnetically controlled. At large distances from the planet, as
determined from equation (\ref{dominate}), the stellar wind pressure
will play an important role.

The star also has a magnetic field that must be taken into account.
If the surface strength of the stellar field is $B_\ast$, the strength
at the location of the planet will be $B \approx B_\ast (R_\ast/a)^3$,
where $a$ is the semimajor axis of the planetary orbit. This
expression would be exact in the limiting case of a pure dipole field
with the planetary orbit in the equatorial plane of the dipole. In
practice, however, the stellar field will be more complicated due to
the stellar wind. For solar-type conditions, the field lines are
closed out to radii $\sim 3 R_\ast$, but they spiral outwards at
larger distances. The planet radius $R_P \sim 10^{10}$ cm is much smaller
than both the stellar radius $R_\ast \sim 10^{11}$ cm and the distance
to the star $a \sim 10^{12}$ cm. As a result, to leading order, the
stellar contribution to the field can be considered to have constant
field strength and constant direction over the region where the
planetary outflow is launched; it is straightforward to show that this
approximation results in error terms of order ${\cal O} (r/a) \sim
10^{-2}$.  At the planetary surface, the ratio of the stellar field
strength to that of the planet is given by
\be
\bstar \equiv {B(a) \over B_P} \approx {B_\ast \over B_P} 
\left( {R_\ast \over a} \right)^3 \, , 
\label{betadef} 
\ee
where $B_\ast$ is the field strength on the stellar surface and $B(a)$
is the stellar field strength evaluated at the position of the planet.
For the regime of parameter space of interest, this field geometry has
an X-point, a location where the magnetic field vanishes (Shu 1992).
Here, the stellar field is (nearly) constant in the vicinity of the
planet, whereas the planetary field decreases as $B \propto 1/r^3$.
To leading order, the radius of the X-point is given by
\be
r_X \approx {a R_P \over R_\ast} 
\left( {B_P \over B_\ast} \right)^{1/3} \sim 10 R_P \, . 
\label{rxapprox}
\ee
For comparison, note that the Hill radius is given by $r_H$ = 
$a (M_P/3M_\ast)^{1/3}$.  For a Jovian planet and solar-type star,
$r_H \approx 0.07 a \sim 7 R_P$. As a result, the sphere of
gravitational influence of the planet (determined by $r_H$) and the
sphere of magnetic influence of the planet (determined by $r_X$) are
approximately the same.

\begin{figure} 
\figurenum{1} 
{\centerline{\epsscale{0.90} \plotone{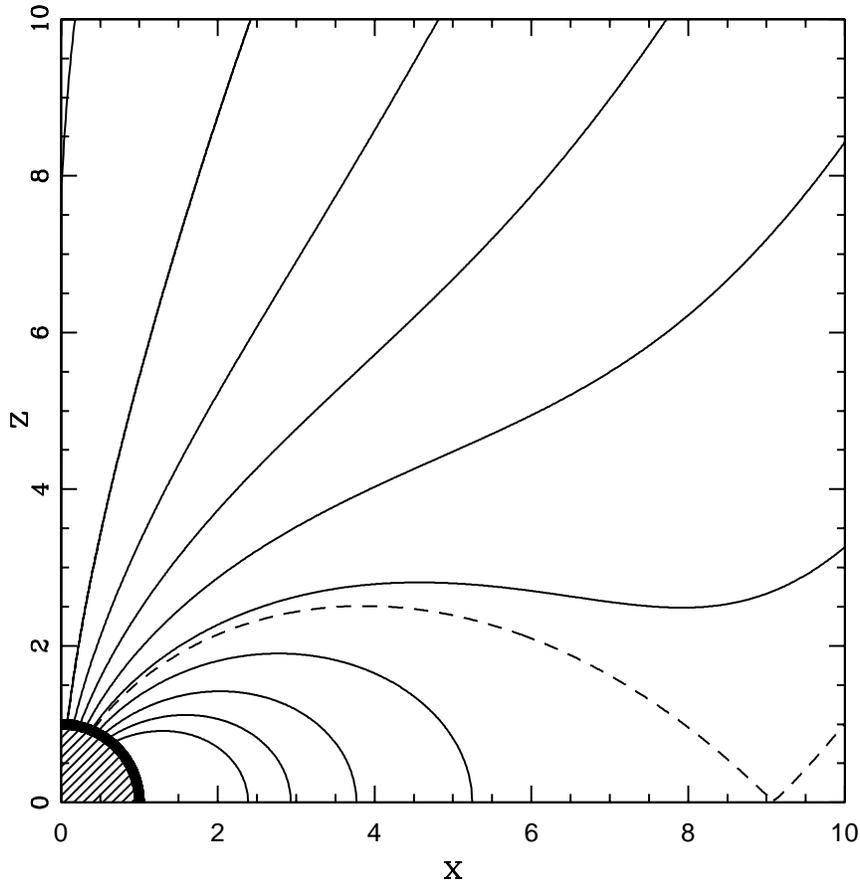} } } 
\figcaption{Magnetic field lines centered on the planet for field 
configurations with background parameter $\bstar \approx$ 0.0014.  
The solid curves show a set of field lines for starting values of
$\sin \theta$ that are evenly spaced. The dashed curve shows the field
line that passes through the X-point. The field lines near the poles
are open, and join onto the background field lines provided by the
star and stellar wind. The field lines near the equator are closed and
end back on the planetary surface. In this coordinate system, 
distances are measured in units of the planet radius, and the star is
located at position $(x,y,z) = (100,0,0)$. }
\label{fig:blines} 
\end{figure}

To illustrate the magnetic field geometries that arise, we plot a
representative collection of magnetic field lines in Figure
\ref{fig:blines}. This plot uses $\bstar = 0.0014$, a value that
occurs (for example) when the star and planet have comparable surface
field strengths and the length scales $R_\ast/R_P = 10$ and $a/R_P$ =
100.  Magnetic field lines that originate from small polar angles on
the planet surface curve off to large distances.  These field lines
will, in general, connect up with stellar field lines, which either
end on the star or are carried out to large distances by the stellar
wind.  Field lines that start at lower latitudes (closer to the
planetary equator) curve back and end on the planetary surface, i.e.,
they are closed.  For a given field configuration, a well-defined
fraction $F_{AP}$ of the planetary surface is exposed by having its
field lines open (relative to the planet).  The field line that passes
through the X-point, shown by the dashed curve in Figure
\ref{fig:blines}, delineates the boundary between the closed and open
field lines.

In principle, currents produced by the flow could modify the planetary
magnetic field, assumed here to have a nearly dipole form. This effect
has been extensively studied, primarily in the context of stellar
winds (e.g., Mestel 1968, Mestel \& Spruit 1987; see also TAL for an
application to extrasolar planet magnetospheres). As shown in the
aforementioned work, currents $J_\perp$ that run perpendicular to the
flow can arise from either vorticity or from the variations of the
Bernoulli constant across field lines. These currents give rise to
perturbations in the magnetic field strength, $B_\perp$, which are of
order
\be
{B_\perp \over B} \sim {4 \pi \rho v^2 \over B^2} \, . 
\label{perpen} 
\ee
As shown above, this ratio is quite small within the sonic surface,
typically less than $10^{-4}$, so that the additional fields produced
by perpendicular currents can be neglected. 

Magnetic fields can also be generated by current sheets, which arise
at the interface between the field lines that carry outflowing
material and those that do not. The difference in flow velocity across
this boundary implies a pressure difference (from the Bernoulli
equation) which results in a perturbation $\delta B$ of the magnetic
field. Applied to the present problem, the results of TAL imply that
$(\delta B)/ B \sim 8 \pi P/B^2$, where $P$ is the thermal pressure
and $B$ is the dipole field of the planet. Over most of the parameter
space of interest, the field produced by current sheets $(\delta B)$
is thus the same order as that produced by perpendicular currents
($B_\perp$, see equation [\ref{perpen}]), and hence can be neglected
to leading order $([\delta B]/B \sim 10^{-4}$). 

Perpendicular currents and current sheets can be neglected when the
pressure $P \ll B^2/(8\pi)$. In the case of stellar winds, this
condition is violated sufficiently far from the stellar surface; at
these large distances currents cannot be ignored, as they lead to open
field lines, which are necessary for winds to escape (Mestel \& Spruit
1987). In the present case, however, an external field allows for open
field lines (relative to the planet) even in the absence of currents.
As shown in Section 3.2, as long as the external field is nonzero,
some of the field lines will be open; the fraction of the planetary
surface that is covered by open field lines is an increasing function
of the external field strength (see equation [\ref{pfraction}]).

The above discussion defines the expected planet radius $R_P$, X-point
radius $r_X$, Hill's radius $r_H$, stellar radius $R_\ast$, and
semimajor axis $a$. The principal calculation of this paper determines
the radius $r_s$ of the sonic point, the location where the outflow is
launched and can escape the planet, and finds $r_s \sim 3 - 4 R_P$
(Section \ref{sec:launch}).  The remaining length scale in the problem
is the scale height $H$ of the planetary atmosphere, where $H$ = $kT/
(\mu g)$ = $kT R_P / (\mu G M_P)$. As a reference point, we evaluate
the scale height at the atmospheric level where most of the stellar
light is absorbed. At this layer, previous work (e.g., MCM) shows that
the effective temperature of the planet $T \sim 1000$ K and hence $H
\sim R_P/100$; the scale height increases to $H \sim R_P/10$ in the
upper atmosphere where the UV photons are absorbed and $T \sim 10^4$
K.  With this specification, the length scales involved in planetary
outflows obey the ordering
\be
H \ll R_P < r_s < r_H \sim r_X \sim R_\ast \ll a \, .  
\ee 
The sphere of gravitational influence of the planet (from the Hill
radius $r_H$) and the sphere of magnetic influence (from the X-point
radius $r_X$) are roughly the same size, and are comparable to the
stellar radius. The sonic surface typically lies at a few planetary
radii and is thus well inside both $r_H$ and $r_X$. On the other hand,
the sphere of influence of the planet is much smaller than the
star-planet distance (the semimajor axis $a$).  This separation of
scales allows for the launch of the outflow to be considered
independently of the subsequent propagation of the flow.

\section{Magnetic Field Geometry with Constant Background} 
\label{sec:field} 

To gain further understanding of this problem, and to simplify the
calculations, we use the method of matched asymptotic expansions.
Specifically, we divide the the problem into two regimes: [1] The
region near the planet where the outflow is launched, and [2] the
region ``far'' from the planet where the magnetic field structure is
determined by the field of the star (modified by the stellar wind).
In the near region, the magnetic field is primarily determined by the
dipole field of the planet, but nonetheless contains a contribution
from the stellar field. However, this stellar contribution is nearly
uniform.  Exploiting this property, we can model the magnetic field
near the planet through the reduced form
\be
{\bf B} = B_P \left[ \xi^{-3} \left( 3 \cos \theta \rhat - \zhat \right) 
+ \bstar \zhat \right] \, , 
\label{reducedfield} 
\ee 
where $\xi = r/R_P$ and where $\bstar$ is defined by equation
(\ref{betadef}). For simplicity we have taken the background field to 
point in the $\zhat$ direction.  Notice that this reduced field is 
axisymmetric, so that the problem becomes two dimensional. In
addition, the field is anti-symmetric with respect to reflections
across the $z$ = 0 plane; since the sign of the field does not affect
the dynamics, the outflow is the same for both hemispheres of the
planet.

In this initial treatment, the dipole field of the planet is augmented
by a constant background field (due to the star). The stellar field is
expected to be dipolar near the stellar surface, but will be modified
by the stellar wind. This stellar wind acts to straighten the field
lines, making them more radial. In any case, the stellar field is
expected to be nearly constant, in both strength and direction, over
the region where the wind is launched.  However, the background field
will not necessarily point in the $\zhat$ direction of the planetary
dipole (as assumed here).  Notice also that the planetary spin, and
hence the planetary dipole, does not necessairly line up with
direction of the orbital angular momentum, i.e., the dipole also has
(in general) an arbitrary direction (e.g., Fabrycky \& Winn 2009).
This paper considers the simplest case where the dipole and the
background field are aligned (equation [\ref{reducedfield}]).  Despite
its simplicity, this ansatz displays the key features of the expected
magnetic configuration: a dipole form near the planet, an
(effectively) straight geometry far from the planet, and some open
field lines even in the absence of currents.  The more general case,
where the two components lie at an arbitrary angle, should be
considered in future work.  In addition, if the star rotates more
slowly than the angular velocity of the planetary orbit, the field
lines will tend to wrap up into spiral configurations (Parker 1958);
this effect should also be included in future studies.

\subsection{Construction of the Coordinate System} 

With this configuration, the magnetic field is current-free and
curl-free, and hence can be written as the gradient of a scalar
field. We define an analogous scalar field $p$ that serves as 
the first field of the coordinate system, i.e., 
\be
p = (\bstar \xi - \xi^{-2}) \cos \theta \, . 
\label{pdef} 
\ee
The gradient $\nabla p$ defines a vector field that points in the
direction of the magnetic field (and hence points in the direction 
of the outflow). Next we construct the perpendicular vector field
$\nabla q$, where the second scalar field $q$ provides the second 
coordinate and is given by
\be
q = \left( \bstar \xi^2 + 2/\xi \right)^{1/2} \sin \theta \, . 
\label{qdef} 
\ee 
The pair $(p,q)$ thus represents a set of perpendicular coordinates in
the poloidal plane, and these can be used instead of the original
spherical coordinates $(\xi, \theta)$ or the cartesian coordinates
$(x,z)$. In this version of the problem, the field is axisymmetric
about the $\zhat$ axis, so we can use the usual azimuthal coordinate
$\phi$ as the third scalar field of the set (the magnetic field is
poloidal, with no toroidal component). Note that both $p$ and $q$ are
dimensionless; one can reinsert factors of the planetary radius $R_P$
(as necessary) to convert back to physical units. Notice also that in
the limit $\bstar \to 0$, one recovers the coordinates for a dipole
(Radoski 1967).

The set of covariant basis vectors $\underline{\epsilon}_j$ arises
from the gradients of the scalar fields that define the coordinates.
If we express these basis vectors in terms of the original spherical
coordinates $(\xi, \theta, \phi)$, the basis takes the form
\be
\epp = (\bstar + 2 \xi^{-3}) \cos \theta \, \rhat - 
(\bstar - \xi^{-3}) \sin \theta \, \thetat \, , 
\ee
\be
\epq = (\bstar + 2 \xi^{-3})^{-1/2} \left[ (\bstar - \xi^{-3})
\sin \theta \, \rhat + (\bstar + 2 \xi^{-3}) 
\cos \theta \, \thetat \, , \right] \, , 
\ee
and 
\be
\epphi = {1 \over \xi \sin\theta} \, \phihat \, . 
\ee 
The quantities $(\rhat, \thetat, \phihat)$ are the usual unit vectors
for spherical coordinates. However, one should keep in mind that the
$\underline{\epsilon}_j$ are basis vectors (not unit vectors), so that
their length is not, in general, equal to unity (for further
discussion, e.g., see Weinreich 1998). In general, the corresponding
scale factors are given by the relation
\be
h_j = \left| {\underline{\epsilon}}_j \right|^{-1} \, , 
\ee
so that the scale factors for this coordinate system can 
be written in the form 
\be
h_p = \left[ (\bstar + 2 \xi^{-3})^2 \cos^2 \theta + 
(\bstar - \xi^{-3})^2 \sin^2 \theta \right]^{-1/2} \, , 
\ee
\be 
h_q = (\bstar + 2 \xi^{-3})^{1/2} 
\left[ (\bstar + 2 \xi^{-3})^2 \cos^2 \theta + 
(\bstar - \xi^{-3})^2 \sin^2 \theta \right]^{-1/2} \, , 
\ee
and 
\be
h_\phi = \xi \sin \theta \, . 
\ee
The general form of the divergence operator is thus given by 
\be
\nabla \cdot {\bf V} = {1 \over h_p h_q h_\phi} 
\left[ {\partial \over \partial p} \left( h_q h_\phi V_p \right) + 
{\partial \over \partial q} \left( h_p h_\phi V_q \right) \right] + 
{1 \over \xi \sin\theta} {\partial V_\phi \over \partial \phi} \, . 
\ee
In this problem, the fields are axisymmetric so that the $\phi$
derivatives vanish. Further, for flow along field lines, the vector
fields (e.g., the velocity field) have only one component and depend
on only one coordinate, so that the divergence operator collapses to
the form 
\be
\nabla \cdot {\bf V} = {1 \over h_p h_q h_\phi} 
{\partial \over \partial p} \left( h_q h_\phi V_p \right) =
{1 \over h_p} {\partial V_p \over \partial p} + 
{V_p \over h_p h_q h_\phi} {\partial \over \partial p} 
\left( h_q h_\phi \right) \, . 
\ee
For convenience, we define the following ancillary functions: 
\be
f \equiv \bstar + 2 \xi^{-3} , \qquad 
g \equiv \bstar - \xi^{-3} , \qquad {\rm and} \qquad 
H \equiv f^2 \cos^2 \theta + g^2 \sin^2 \theta \, . 
\label{ancillary} 
\ee
In terms of these functions, we note that $(h_q h_\phi)$ = 
$q H^{-1/2}$ and that $|{\bf B}|^2$ = $H B_P^2$. 

This specification of the divergence operator is implicit. One could
invert equations (\ref{pdef}) and (\ref{qdef}), and then write the
spherical coordinates $(\xi,\theta)$, the scale factors $(h_p, h_q,
h_\phi)$, and the ancillary functions $(f,g,H)$ as functions of the
new coordinates $(p,q)$. However, the definitions of $(p,q)$ are cubic
functions of $(\xi,\theta)$, so that the solutions of the cubic
inversion are complicated and unwieldy (but still can be written down
analytically). For clarity, we leave this construction in implicit
form.

\subsection{Fractional Active Area of the Planetary Surface} 

With the field configuration of equation (\ref{reducedfield}), some
fraction $F_{AP}$ of the magnetic field lines that originate on the
planetary surface are open, whereas some field lines curve back onto
the planet. Using this simplified model for the magnetic field, we can
determine the fraction $F_{AP}$. In this context, the open field lines
that continue to ``infinity'' in this reduced problem join onto the
magnetic field lines of the star and stellar wind (although the
subsequent curvature of the field lines is not captured by this
model).  One can define a benchmark mass outflow rate to be that
obtained for spherical flow over the entire planetary surface. We
note that the fraction $F_{AP}$ of the surface that supports open
field lines is related to --- but is not equivalent to --- the
fraction of this benchmark outflow rate that the planet produces 
(see Section 4.4). 

The first step is to solve for the magnetic field lines. For this
field configuration, this construction can be done analytically.
Assume that a field line begins at coordinates $(\xi = 1, \theta_0)$
corresponding to the planetary surface. The field line can then be
represented as a curve in the plane such that 
\be
\left( {\sin\theta \over \sin \theta_0} \right)^2 = \xi 
{2 + \bstar \over 2 + \bstar \xi^3} \, . 
\label{fieldlinesimple} 
\ee
Note that this expression is equivalent to the statement $q$ = 
$\sin \theta_0 (2 + \bstar)^{1/2}$ = {\sl constant}; this result must
hold since the coordinate $q$ was constructed so that $\nabla q$ is
perpendicular to the field lines.  Using this solution, we can solve
for the angular coordinate $\theta_X$ for the field line that goes
through the X-point. In this case, the X-point radius $\xi_X$ is given
by the condition
\be 
\bstar \xi_X^3 = 1 \, . 
\ee
Note that this result is exact for the reduced field configuration
considered here. The critical angle $\theta_X$ is given by
\be
\sin^2 \theta_X = 3 \bstar^{1/3} / (2 + \bstar) \, , 
\label{thetax} 
\ee
where this expression is valid for $\bstar \le 1$.  The field lines
that originate at small angles, $0 \le \theta \le \theta_X$, with
$\theta_X$ defined above, are those that are open (reach spatial
infinity) in this reduced problem. In the full problem, these open
field lines join onto the background field lines of the star and
stellar wind. The resulting fraction $F_{AP}$ of the surface area 
of the planet that can support an outflow is then given by 
\be
F_{AP} = 1 - \left[ 1 - {3 \bstar^{1/3} \over 2 + \bstar} 
\right]^{1/2} \, . 
\label{pfraction} 
\ee
Note that for $\bstar \ge 1$, the background (stellar) field dominates
that of the planet, field lines originating from all planetary
latitudes are open and hence $F_{AP} = 1$. In the opposite limit of
small $\bstar \ll 1$, the fraction $F_{AP} \approx 0.75 \bstar^{1/3}$;
since $\bstar \sim 10^{-3}$ for typical cases, only about 10 percent of
the planet surface can support outflow.

With this field geometry, the field lines become asymptotically
straight, so that the outflow has a cylindrical form in the limit 
$\xi \to \infty$.  The outer surface of the resulting outflow cavity
is delineated by the field line that passes through the X-point. If we
use equation (\ref{fieldlinesimple}) for the critical streamline, and
then take the limit $\xi \to \infty$, we find that the radius
$\varpi_\infty$ of this cylinder is given by 
\be
\varpi_\infty = \sqrt{3} \, r_X = \sqrt{3} \, \bstar^{-1/3} R_P \, 
= \sqrt{3} \, (B_P/B_\ast)^{1/3} \, \left( a R_P / R_\ast \right) \, . 
\label{infradius} 
\ee
This cylindrical flow represents the outer limit of the inner problem,
the regime where the outflow is launched from the planetary surface.
This flow also represents the inner limit of the outer problem, the
regime where the flow follows the stellar field lines.  Note that for
typical parameters $\varpi_\infty \approx$ 10 $R_P \approx R_\ast$.

Since the magnetic field strength vanishes along the critical
streamline, the magnetic pressure will not be strong enough to
dominate the ram pressure near the X-point. We can estimate the
fraction of streamlines that are affected by this issue: Consider the
surface where $\xi = \xi_X = \bstar^{-1/3}$. For a given angle $\theta$
on this surface, the magnetic field strength (from equation
[\ref{reducedfield}]) is given by $B = 3 B_P \bstar \cos \theta$.  For
the flow to remain magnetically controlled, the field strength must be
larger than a critical value $B_C$ determined by the ram pressure of
the outflow ($B_C^2 = 8 \pi \rho v^2$), where we expect $B_C \approx$
0.001 G for typical cases (see equation [\ref{dominate}]). This
condition requires $\cos \theta > B_C / (3 B_P \bstar)$, which in turn
restricts the coordinate $q$ that specifies the streamlines to the range
defined by
\be
q^2 < \left[ 1 - \left( {B_C \over 3 B_P \bstar} \right)^2
\right] q_X^2 \, . 
\ee
Since we expect $B_C \sim \bstar B_P$, the range of streamlines that
are unaffected by the X-point issue is roughly given by $q < q_X
\sqrt{8}/3$. In other words, for $\sim$94 percent of the streamlines,
the flow remains magnetically controlled over the entire space. For
the remaining $\sim$6 percent, the flow is magnetically controlled except
for a small region (of size $(\Delta L) \sim r_X/3$) surrounding the
X-point.  Note that outflow can still take place along these
streamlines, but that the flow pattern will be slightly different than
that given by the unperturbed magnetic field geometry.  For the
remainder of this paper, for simplicity, we consider the flow to take
place over the full range of open streamlines $q \le q_X$, while
noting that this approximation may overestimate the outflow rates by a
few percent.

\section{Outflows from Planetary Surfaces} 
\label{sec:launch} 

Given the specification of the magnetic field structure and hence the
flow geometry (Sections \ref{sec:overview} and \ref{sec:field}),
we now consider the launch of a wind or outflow from the surface of
the planet. After writing down the full set of equations of motion, we
consider a reduced version of the problem where the flow is taken to
be isothermal. For this case, the solutions for the dimensionless
fluid fields can be found analytically, including the required
conditions for the flow to pass smoothly through the sonic transition.
In order to complete the solution, we must then specify the values for
the physical parameters, i.e., the density $\rho_1$ at the inner
boundary and the sound speed $a_s$ (taken to be constant).

\subsection{Formulation of the Wind/Outflow Problem} 

The equations of motion for this problem include the  
continuity equation, 
\be
{\partial \rho \over \partial t} + \nabla \cdot \left( 
\rho \uvec \right) = 0 \, ,  
\label{continue} 
\ee
the force equation, 
\be
{\partial \uvec \over \partial t} + \uvec\cdot \nabla \uvec = 
- \nabla \Psi - {1\over \rho} \nabla P + {1 \over 4 \pi \rho} 
\left( \nabla \times \bvec \right) \times \bvec \, , 
\label{force} 
\ee
the energy equation 
\be
\rho \left( {\partial \energy \over \partial t} + 
\uvec \cdot \nabla \energy \right) = - P \nabla \cdot \uvec
+ \Gamma - \Lambda \, , 
\label{energy} 
\ee
and the evolution equation for the magnetic field 
\be
{\partial \bvec \over \partial t} + \nabla \times 
\left( \bvec \times \uvec \right) = - \nabla \times 
\left( \eta_{\rm res} \nabla \times \bvec \right) \, , 
\label{induction} 
\ee 
where $\eta_{\rm res}$ is the resistivity. This paper considers the
magnetic fields to be fixed, and strong enough to not be changed by
the outflow. In a full treatment, however, the back reaction of the
outflow on the magnetic field should be taken into account.

In this paper, we consider the gravitational potential $\Psi$ to be
that of the planet, which is taken to be spherical with mass $M_P$ and
radius $R_P$. Since the planet orbits the star, the full potential has
an additional contribution from the rotating frame of reference. The
order of this correction term is ${\cal O}( M_\ast r^3 / M_P a^3)$, so
that it has has size $\sim 10^{-3}$ near the planet surface and size
$\sim 0.03$ near the sonic surface. As a result, this term does not
greatly affect the launch of the outflow and is not included here. To
consistent order, we also ignore the tidal forces from the stellar
gravitational field (note that the sonic surface is well inside the
Roche radius).  We thus work in the regime where the gravitational
force is dominated by that of the planet.

In the energy equation (\ref{energy}), $\energy$ is the specific
energy of the fluid, $\Gamma$ is the heating rate (per unit volume),
and $\Lambda$ is the cooling rate. The heating is primarily due to UV
flux from the central star (e.g., see MCM and references therein) so 
that $\Gamma$ can be written in the form
\be
\Gamma = \effabsorb F_{UV} e^{-\tau} \sigmauv n_0 \, , 
\label{heating} 
\ee 
where $\effabsorb$ is the fraction of the UV energy that is deposited
by heat, $\sigmauv$ is the cross section for UV photons, $F_{UV}$ is
the unattenuated UV flux of the star at the location of the planet,
and $\tau$ is the optical depth (from the star to the point where the
heating term is evaluated).

The cooling process is primarily due to Lyman-$\alpha$ radiation that
is emitted by Hydrogen atoms as they are excited via collisions (Black
1981). To leading order, the cooling term can be represented by a
function of the form
\be
\Lambda \approx C n_0 n_+ e^{-T_C/T} \, , 
\label{cooling} 
\ee 
where $C$ = $7.5 \times 10^{-19}$, $T_C$ = 118348 K; the resulting
cooling rate has units of erg cm$^{-3}$ s$^{-1}$.  Note that this
particular form is only valid for temperatures $T < 12,000$ K; at
higher temperatures the gas cools even more efficiently (Spitzer
1978).  In any case, the cooling rate is large enough that the gas
temperature never increases beyond an effective maximum $T_{\rm max}
\approx 10^4$ K. In order for Lyman-$\alpha$ radiation to act as the
primary cooling mechanism, it must dominate over other processes and
the radiation must be able to escape; both of these conditions are
met, as shown in the Appendices of MCM.

Since the heating and cooling rates depend on the state of
ionization, through the number densities of the neutrals $n_0$ and
ions $n_{+}$, we also need the equation of ionization balance:
\be
{\partial n_{+} \over \partial t} + \nabla \cdot \left( \uvec n_+ \right) 
= {F_{UV} \over \langle h \nu \rangle } e^{-\tau} \sigmauv n_0 
- \alpha_R n_+^2 \, ,
\label{ionbalance} 
\ee
where the recombination coefficient 
$\alpha_R \approx 2.7 \times 10^{-13} (T/10^4 {\rm K})^{-0.9}$ cm$^3$
s$^{-1}$ (Storey \& Hummer 1995).  For the remaining parameters, 
we take $\sigmauv$ = $2 \times 10^{-18}$ cm$^2$ and 
$\langle h\nu \rangle$ = $2.2 \times 10^{11}$ erg 
(e.g., Spitzer 1978).

Note that the planet is only heated on the side facing the star,
whereas cooling takes place over the entire surface of the planet.
This work implicitly assumes that the UV heating is distributed
uniformly throughout the upper atmosphere of the planet. This
approximation, in turn, is valid when zonal winds are strong enough to
provide the required redistribution.  Although the issue is not
settled, current models suggest that strong winds are present, so that
uniform heating is a reasonable approximation (for further discussion,
e.g., see Batygin \& Stevenson 2009, Langton \& Laughlin 2008, and
references therein).

As formulated here, equations (\ref{continue} -- \ref{ionbalance}) make
up a compete set that can be solved for the fluid fields. However,
even if we use a simplified magnetic field geometry, with the
coordinates constructed in Section \ref{sec:field}, the problem
remains intrinsically three dimensional: Although the flow is
axisymmetric and follows the coordinate $p$ defined by equation
(\ref{pdef}), the heating comes from the central star which lies off
to one side.  As a result, the incoming photons do not follow the
coordinates, and a full solution for the heating/cooling of the
outflow requires one to solve a three-dimensional radiative transfer
problem. Before embarking on that task, it is useful to have solutions
for an approximate treatment. Toward this end, a simplified version of
the problem is formulated in the next section.

\subsection{Reduced Equations of Motion} 

In this section, we consider steady-state solutions and assume that the 
magnetic field structure due to the planet (and the star) are strong
enough to dominate the flow. As a result, in this regime, the magnetic
field is fixed and current-free. The continuity, force, and induction 
equations thus reduce to the forms
\be
\nabla \cdot \left( \rho \uvec \right) = 0 \, , \qquad 
\uvec\cdot \nabla \uvec + \nabla \Psi + {1\over \rho} \nabla P = 0 \, , 
\qquad {\rm and} \qquad \bvec = \Upsilon \rho \uvec \, , 
\label{simple} 
\ee
where the parameter $\Upsilon$ is constant along streamlines (e.g.,
Shu et al. 1994, Cai et al. 2008).

The velocity vector ${\bf u}$ follows the magnetic field lines, which
follow the coordinate $p$ in the system constructed in Section
\ref{sec:field}. In other words, the flow velocity has only one
component, which points in the direction of the magnetic field 
$\hat p$ = $h_p \epp$ (by construction).  Next we assume that the flow
is isothermal with constant sound speed $a_s$ and define the following
dimensionless quantities
\be
u = u(p) \equiv {|{\bf u}| \over a_s} , \qquad 
\alpha \equiv {\rho \over \rho_1} , \qquad 
\xi \equiv {r \over R_P} , \qquad {\rm and} \qquad 
\psi \equiv {\Psi \over a_s^2} \, . 
\ee 
Here, $R_P$ is the radius of the planet and $\rho_1$ is the density at
the inner boundary $\xi$ = 1.  The continuity equation thus takes the form
\be
\alpha {\partial u \over \partial p} + 
u {\partial \alpha \over \partial p} = - 
{\alpha u \over h_q h_\phi} {\partial \over \partial p} 
\left( h_q h_\phi \right) \, , 
\ee
and the force equation becomes 
\be
u {\partial u \over \partial p} + {1 \over \alpha} 
{\partial \alpha \over \partial p} = - {\partial \psi \over \partial p} 
= - {\partial \psi \over \partial \xi} {\partial \xi \over \partial p} \, . 
\ee
These equations can be integrated immediately to obtain the solutions 
\be
\alpha u h_q h_\phi = \alpha u q H^{-1/2} = \lambda \, , 
\label{simplecont} 
\ee
and 
\be
{1 \over 2} u^2 + \log \alpha + \psi = \varepsilon \, . 
\label{simpleforce} 
\ee
Although the potential $\psi$, in general, contains additional
contributions (e.g., tidal forces), we specialize to the case that
includes only the gravitational potential of the planet so that $\psi$
= $-b/\xi$, where $b \equiv G M_P /(a_s^2 R_P)$.  Note that the
quantity $h_q h_\phi$ = $q H^{-1/2}$ is proportional to the inverse of
the magnetic field strength (consistent with the third part of
equation [\ref{simple}]).  The parameters $\lambda$ and $\varepsilon$
are constant along streamlines, but are not, in general, the same for
all streamlines (they are functions of $q$).  In order for the flow to
pass smoothly through the sonic point, only particular values of the
constant $\lambda$ are allowed. This constraint is considered in the
following section.

The boundary conditions at the planetary surface take the form
\be
\xi = 1 \, , \qquad \alpha = 1 \, , \qquad {\rm and} \qquad
u = u_1 = \lambda H_1^{1/2} / q \, . 
\label{innerbc} 
\ee
Since $\lambda$ is determined by the conditions at the sonic point,
$u_1$ is specified. In addition, the remaining parameter $\varepsilon$
is determined by evaluating the force equation at the inner boundary,
i.e., 
\be
\varepsilon = {1 \over 2} u_1^2 - b = {\lambda^2 H_1 \over 2 q^2} - b \, . 
\label{setenergy}
\ee
The outflow starts with subsonic speeds so that $u_1 \ll 1$ (below we
find that $u_1 \sim \lambda/q \sim 0.01$), whereas typical planet
properties imply that $b \sim 10$.  As a result, one can use the
approximation $\varepsilon \approx - b$ with good accuracy. 

\subsection{Sonic Point Conditions} 

Critical points in the flow arise when the fluid speed is equal to the
transport speed. In general, magnetic media support three types of MHD
waves and hence allow for three types of critical points (e.g., Shu
1992). In this case, however, the flow is confined to follow the
magnetic field lines, so that only one possible critical point arises,
in this case where the flow speed equals the sound speed.  For the
equations of motion (\ref{simplecont}) and (\ref{simpleforce}), the
required matching conditions at the sonic point take the form
\be
u^2 = 1 \qquad {\rm and} \qquad 
{1 \over h_q h_\phi} {\partial \over \partial p} 
\left( h_q h_\phi \right) = {\partial \Psi \over \partial \xi} 
{\partial \xi \over \partial p} = {b \over \xi^2} 
{\partial \xi \over \partial p} \, . 
\ee
We must thus evaluate the geometrical factor $\cal G$ defined by
\be
{\cal G} \equiv {1 \over h_q h_\phi} {\partial \over \partial p} 
\left( h_q h_\phi \right) \, . 
\label{gdef} 
\ee
Note that in spherical coordinates, this factor would have the usual
form $2/\xi$.  In terms of the ancillary functions from equation
(\ref{ancillary}), the partial derivatives can be expressed as
\be
{\partial \xi \over \partial p} = {f \cos \theta \over H} 
\qquad {\rm and} \qquad {\partial \theta \over \partial p} 
= - {g \cos \theta \over \xi H} \, , 
\label{partialderiv} 
\ee
and the geometrical factor $\cal G$ takes the form  
\be
{\cal G} = {3 \cos \theta \over H^2 \xi^4} 
\left[ 2 f^2 \cos^2\theta - g (g + 2 f) \sin^2\theta \right] \, . 
\ee
The matching condition at the sonic point can then be 
written in the form  
\be
b = {3 \over f H \xi^2} \left[ 2 f^2 \cos^2\theta - 
g (g + 2f) \sin^2\theta \right] \, . 
\label{matching}
\ee
One can eliminate the explicit angular dependence from this expression
using the result $\sin^2 \theta = q^2 / \xi^2 f$, and noting that $q$
is a constant along the direction of the flow.  The right hand side of
the above equation (for given $q$) thus becomes a function of $\xi$ only, 
\be
{b \over 3} = {2 f^2 - \left( g^2/f + 2g + 2f \right) q^2 / \xi^2 
\over f^3 \xi^2 + \left(g^2 - f^2 \right) q^2 } \, . 
\label{matchingtwo}
\ee
For given planetary properties (set by the value of $b$), magnetic
field strength ratio (set by $\bstar$), and starting angle $\theta_0$
of the streamline (set by $q$), equation (\ref{matchingtwo}) provides
an algebraic expression that can be solved for the value of $\xi$ =
$\xi_s$ at the sonic point. With $\xi_s$ specified, the angle
$\theta(\xi_s)$ is also specified, and hence the value of $p = \xi g
\cos\theta$ is determined. Finally, the value of the parameter
$\lambda$ that allows for smooth flow through the sonic point is given by
\be
\lambda = \lambda(q) = q H_s^{-1/2} \exp \left[
{\lambda^2 H_1 \over 2 q^2} + {b \over \xi_s} - 
b - {1\over2} \right] \, , 
\label{lambda} 
\ee
where the subscript `s' (`1') implies that the quantity is evaluated
at the sonic point (inner boundary). Equation (\ref{lambda}) provides
an implicit solution for the parameter $\lambda$.  However, the
$\lambda^2$ term on the right hand side of equation (\ref{lambda}) is
extremely small (it is equal to $u_1^2/2 \ll 1$) and can be ignored to
leading order; doing so results in a direct expression for the
parameter $\lambda$ (after equation [\ref{matchingtwo}]) has been
solved to find the value of $\xi_s$). We also note that one can define
an alternate parameter ${\tilde \lambda} \equiv \lambda/q$, which is 
useful because it is easier to find solutions for ${\tilde \lambda}$
(e.g., when $q \to 0$). 

\begin{figure} 
\figurenum{2} 
{\centerline{\epsscale{0.90} \plotone{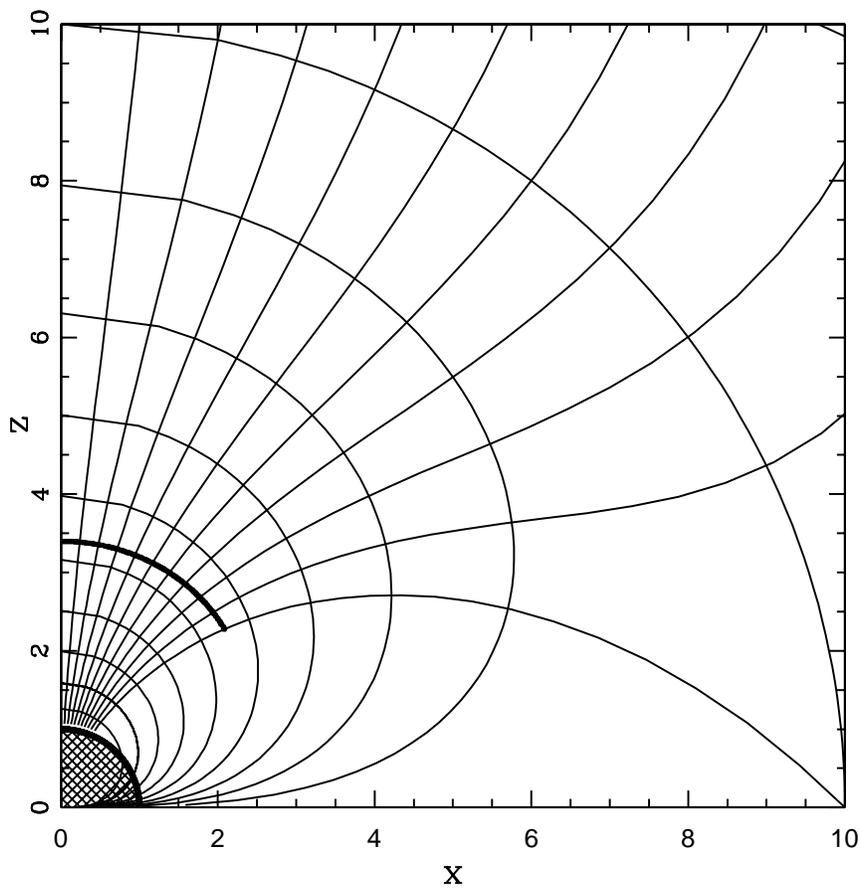} }} 
\figcaption{Sonic transition surface for planet with dimensionless 
gravitational potential depth $b$ = $G M_P / (a_s^2 R_P)$ = 10 and
background magnetic field strength parameter $\bstar$ = $10^{-3}$. The
heavy solid curve near $\xi$ = 3.3 shows the location where the flow
passes through the sonic point. The light solid curves show the
underlying $(p,q)$ coordinate system. The cartesian coordinates 
$x$ and $z$ are given in units of the planetary radius. } 
\label{fig:sonicsurf} 
\end{figure}

Figure \ref{fig:sonicsurf} shows the sonic surface for a planet/star
system with typical properties. Here, the parameter $\bstar = 10^{-3}$,
which holds when the surface fields on the planet and star are equal,
and the semimajor axis $a$ = 10 $R_\ast$; the parameter $b$ = 10,
which holds for planets with mass $M_P = 0.75 M_J$, radius $R_P$ =
$10^{10}$ cm, and sound speed $a_s$ = 10 km/s.  The sonic transition
occurs at nearly constant dimensionless radius ($\xi \approx 3.3$), as
delineated by the heavy curve. The light solid curves show the
underlying coordinate system, i.e., the lines of constant $p$ and
$q$. Only the streamlines (constant $q$) corresponding to open
magnetic field lines are shown. The last coordinate line shown is the
one passing through the X-point. For the perpendicular coordinate, the
value $p$ = 0 corresponds to the surface passing through the X-point,
and that surface is a sphere.  Negative values of $p$ correspond to
surfaces that are more highly curved (compared to a sphere), whereas
positive values of $p$ produce flatter surfaces. Note that the sonic
surface lies well within the surface passing through the X-point.  As
a result, the flow diverges significantly faster than that of a
spherically symmetric wind.

\begin{figure} 
\figurenum{3} {\centerline{\epsscale{0.90} \plotone{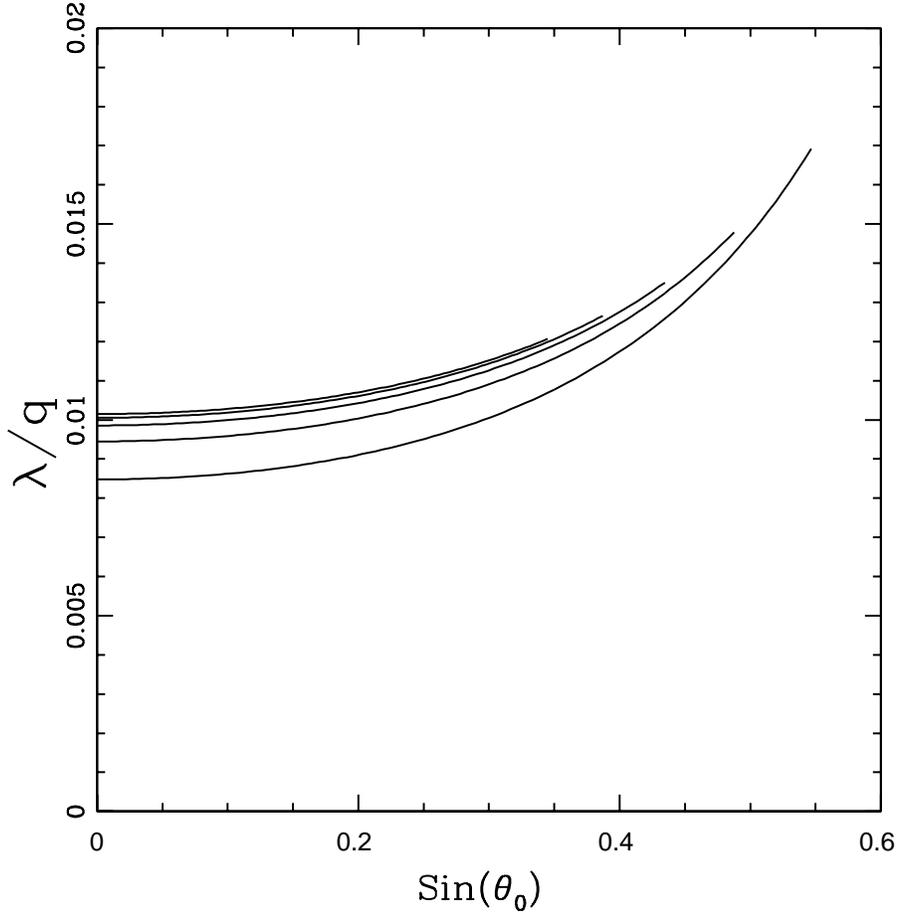} }}
\figcaption{Values of the parameter $\lambda/q$ that allow the outflow
to pass smoothly through the sonic point. The parameters $\lambda$ are
shown here as a function of $\sin \theta_0$ = $q (2 + \bstar)^{-1/2}$,
which determines the starting point of the streamlines. All cases use
$b$ = 10 (where $b$ = $G M_P / (a_s^2 R_P)$ sets the depth of the
gravitational potential of the planet). Curves are shown for a range
of the parameter that sets the relative strength of the stellar and
planetary magnetic fields: $\bstar$ = 0.008 (bottom curve), 0.004,
0.002, 0.001, and 0.0005 (top curve). For smaller values of $\bstar$,
the curves converge toward a well-defined locus (although the range in
$\sin \theta_0$ shrinks). }
\label{fig:lambda} 
\end{figure}

The right hand side of equation (\ref{matchingtwo}) is not, in
general, a monotonic function of $\xi$; it reaches a maximum and then
decreases in the limit $\xi \to \infty$. If the maximum value is too
small, then equation (\ref{matchingtwo}) has no solutions, and the
outflow will not pass through the sonic point within the context of
this simplified model. One can show that the right hand side of
equation (\ref{matchingtwo}) is a increasing function of $q$ for
sufficiently small $\xi$ (in the regime where matching occurs). As a
result, the minimum value occurs for the smallest value of $q$, i.e.,
along the pole where $q$ = 0. For this streamline, the matching
condition is the most difficult to meet; the sonic point is thus given
by solutions to the cubic polynomial $b = 6 / f \xi^2$, or,
equivalently,
\be
\bstar \xi^2 + 2/\xi = 6/b \, .  
\label{failpoint} 
\ee 
This equation has no real solutions if either parameter $\bstar$ or $b$
is too large; solutions require $\bstar b^3 \le 8$, or equivalently 
$\bstar \le \bstar_{\rm max}$ = $8/b^3$.  Since $b \sim 10$, the value of
$\bstar_{\rm max} \sim 0.01$. Only smaller values of 
$\bstar < \bstar_{\rm max}$ allow for smooth outflow solutions.

For solutions that pass through the sonic point, so that $\lambda$ is
specified, Figure \ref{fig:lambda} presents the resulting values of
$\lambda/q$ over the range of allowed streamlines. These results are
shown for a range of the magnetic field strength parameter $\bstar$.
For the planetary properties used here, where $b$ = 10, the maximum
allowed value of $\bstar$ = 0.008 is shown by the lowest curve in
Figure \ref{fig:lambda}. Larger values of $\bstar$ do not allow for a
sonic transition over the full range of open field lines.  The
resulting $\lambda/q$ for smaller values of $\bstar$ converge toward
the result for $\bstar = 0$. In addition, the range of allowed
streamlines, given here by the range of allowed starting angles
$\theta_0$, decreases with decreasing $\bstar$ (see equation
[\ref{thetax}]). Notice also that $\lambda/q$ is relatively slowly
varying over the range of allowed streamlines. For the largest
possible $\bstar$, $\lambda/q$ varies by a factor of $\sim2$ from the
pole to the angle of the last open field line; the results for 
smaller $\bstar$ show much less variation.

Note that the results derived above make sense in the limits: For flow
near the planetary surface, where the magnetic field is determined by 
the dipole of the planet, the sonic point condition of equation
(\ref{matchingtwo}) reduces to the form $b = 3 \xi$ (along the pole).
This solution results from the continuity equation $\partial_\xi
(\xi^3 \rho u) = 0$, where this form for the divergence operator is
expected for flow that follows a dipole field. In the opposite limit
where $\xi \to \infty$, the field lines and the streamlines become
asymptotically straight and point in the $\zhat$ direction. In this
limit, we recover the results for one-dimensional flow, where the
continuity equation has the form $\partial_z (\rho u)$ = 0. The
solutions that smoothly pass through the sonic point are those where
the magnetic field of the planet dominates, so that the sonic point
$\xi_s \approx b/3$. Using this result, we can define a fiducial value
for the parameter $\lambda/q$, i.e., 
\be
\left( \lambda/q \right)_0 =  {b^3 \over 54} \exp 
\left[ 5/2 - b \right] \approx \, 0.2256 \, \, b^3 {\rm e}^{-b} \, . 
\label{lambdafidu}
\ee
For expected planetary properties with $b \approx 10$, we find
$(\lambda/q)_0 \approx 0.01$. Notice also that $\lambda/q$ sets the
scale for the mass outflow rate and that this quantity decreases
exponentially with increasing depth of the gravitational potential
well of the planet (set by $b$).

For solutions that smoothly pass through the sonic transition, we can
also find the asymptotic speed $u_\infty$, i.e., the value realized in
the limit $\xi \to \infty$. In this regime, the reduced equations of 
motion (\ref{simplecont}) and (\ref{simpleforce}) are simplified
further to take the forms 
\be
u \alpha = \lambda \bstar / q \qquad {\rm and} \qquad 
{1 \over 2} u^2 + \log \alpha = \varepsilon \, ,
\ee
where both $\lambda$ and $\varepsilon$ are known functions of the
parameter $q$ that specifies the streamline.  The asymptotic speed 
$u = u_\infty$ is thus given by the expression 
\be
u_\infty^2 - \log u_\infty^2 = 2 \varepsilon + 
2 \log(q / \bstar \lambda) \, .
\label{uinf} 
\ee
Note that equation (\ref{uinf}) has two roots whenever real solutions
exist, which requires the right hand side to be greater than unity.
One root corresponds to $u_\infty^2 < 1$, whereas the (physical) root
of interest corresponds to $u_\infty^2 > 1$.  

\begin{figure} 
\figurenum{4} 
{\centerline{\epsscale{0.90} \plotone{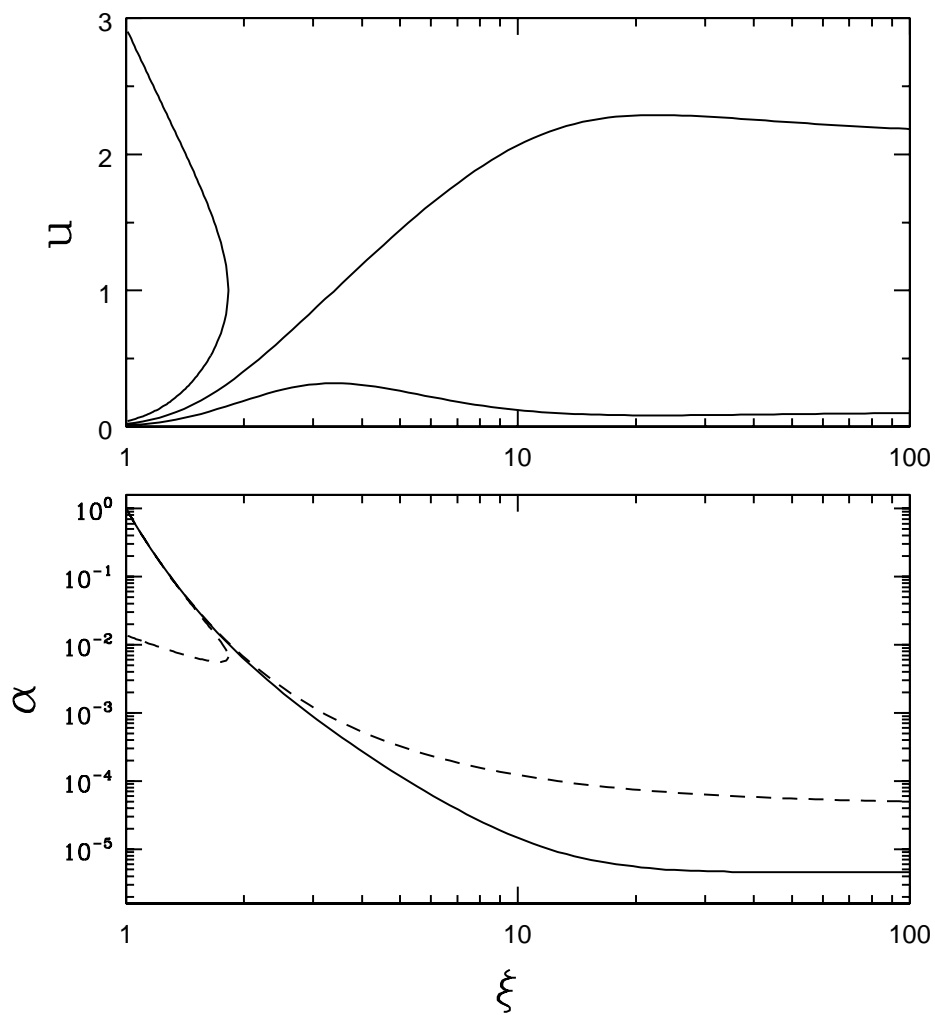} }}
\figcaption{Top panel shows the velocity profiles for outflows from 
planetary surfaces for flow with $q$ = 0 (from the planetary pole).
The three curves show the velocity as a function of radius $\xi$ for
the critical value of $\lambda/q$ (central curve), for a subcritical
case (bottom curve) and a supercritical case (top curve). These
velocity profiles were calculated using $b$ = 10 and $\bstar =
10^{-3}$. Bottom panel shows the corresponding density profiles.
The critical case is shown as the solid curve, whereas the dashed 
curves show the subcritical case (which continues to large $\xi$) 
and the supercritical case (which is confined to small $\xi$). } 
\label{fig:schematic} 
\end{figure}

Figure \ref{fig:schematic} shows the dimensionless velocity and
density profiles for outflows characterized by different values of the
constant $\lambda$. These profiles are shown for flow from the poles,
so that $q$ = 0 and the flow direction is given by $\rhat = \zhat$.
In the top panel, the central curve shows the velocity profile for the
critical value of $\lambda/q$ that allows the flow to pass smoothly
through the sonic point.  If the value of $\lambda/q$ is too small
(bottom curve), then the outflow speed never reaches the sound speed.
Instead, the flow velocity reaches a maximum at the radius of the
sonic point and then decreases for larger $\xi$. On the other hand, if
the value of $\lambda/q$ is too large (top curve), the outflow cannot
reach the sonic point in a smooth manner. This plot is thus analogous
to that found for the well-known Parker wind (see Figure 1 of Parker
1965) and for Bondi-Hoyle accretion (e.g., see Shu 1992 for further
discussion).

The bottom panel of Figure \ref{fig:schematic} shows the corresponding
density profiles. Here the solid curve shows the result for the
critical value of $\lambda/q$ that allows for a smooth sonic
transition. The supercritical and subcritical cases are shown as the
dashed curves. Note that the density profile is extremely steep near
the planetary surface, and then levels out at large radii $\xi$. This
behavior is a reflection of the divergence operator, which follows the
magnetic field lines, which in turn spread rapidly near the planet and
become straight at large distances $\xi$. Notice also that the density
for the subcritical case is larger than that of the critical case for
$\xi \gg 1$. Although the mass outflow rate is smaller for the
subcritical case (which lowers the density), the asymptotic speed (see
the top panel) is much smaller and this latter effect increases the
density and produces the observed behavior. 

For completeness, we can find the values of the parameter $\lambda/q$
for field configurations where the matching condition (equation
[\ref{matching}]) at the sonic point cannot be met. In this case, 
the largest value of $\lambda$, and hence the largest outflow rate, 
occurs when the outflow velocity approaches the sound speed in 
the limit $\xi \to \infty$. In this case, we can combine equations 
(\ref{simplecont}) -- (\ref{setenergy}), while setting $u \to 1$, 
and then solve for the value of the parameter $\lambda/q$: 
\be
\lambda/q = \bstar^{-1} \exp \left[ - (b + 1/2) + 
{\lambda^2 H_1 \over 2 q^2} \right] \approx 
\bstar^{-1} \exp \left[ - (b + 1/2) \right] \, . 
\label{lambdafail} 
\ee

\subsection{Dimensionless Mass Outflow Rate} 

In this geometry, the continuity equation reduces to the form 
$\alpha u q H^{-1/2}$ = $\lambda$ = {\sl constant} (see equation
[\ref{simplecont}]).  Because the quantity $q H^{-1/2}$ does not have
units of area, the constant $\lambda$ is not the mass outflow rate.
As a result, we need to find the relationship between the outflow rate
and the quantities that appear in the equations of motion.

In the limit of large $\xi \to \infty$, the dimensionless mass outflow 
rate ${\dot m}_\infty$ is given by the integral 
\be
{\dot m}_\infty = 2 \int_0^{{\tilde \varpi}_\infty} 
2 \pi {\tilde \varpi} d {\tilde \varpi} \, 
\left( \alpha u \right)_\infty \, , 
\label{mdotfirst} 
\ee 
where ${\tilde \varpi} = \varpi/R_P$ and where the dimensionless outer
radius of the outflow ${\tilde \varpi}_\infty$ can be determined from
equation (\ref{infradius}). The leading factor of two arises because
the wind flows from both the northern and southern hemispheres of the
planet.  From equation (\ref{simplecont}) we find that 
$(\alpha u)_\infty$ = $H^{1/2} u_1 H_1^{-1/2}$, where the subscripts
`1' indicate that the quantities are evaluated at the inner boundary. 
In the limit $\xi \to \infty$, $f \to \bstar$, $g \to \bstar$, and 
$H \to \bstar^2$.  Using the streamline equation, the cylindrical
radius ${\tilde \varpi}$ is related to the orthogonal coordinate $q$
through the expression ${\tilde \varpi}^2 = q^2/\bstar$. After changing
variables, the dimensionless outflow rate takes the form
\be
{\dot m}_\infty = 4 \pi \int_0^{q_X} u_1 \, q \, dq \, H_1^{-1/2} \, , 
\label{largemdotint} 
\ee
where $H_1$ is a function of $q$ and where $q_X^2 = 3 \bstar^{1/3}$. 
In general, the starting speed $u_1$ will not be the same for all 
streamlines, i.e., it will be a function of $q$. If we replace $u_1$ 
in the integral by the appropriate mean value $\langle u_1 \rangle$, 
equation (\ref{largemdotint}) can be evaluated to find 
\be
{\dot m}_\infty = 
{4 \pi \langle u_1 \rangle (2+\bstar)^2 \over 3 (1 + 2 \bstar)} 
\left\{ 1 - \left[ 1 - {9 \bstar^{1/3} (1 + 2 \bstar) 
\over (2 + \bstar)^3 } \right]^{1/2} \right\} \, .  
\label{mdotexpress} 
\ee
This expression is valid for $\bstar \le 1$. For larger values of
$\bstar$, the stellar contribution dominates the magnetic field of the
planet over its entire surface, and the dimensionless mass outflow
rate reduces to the spherically symmetric form ${\dot m} = 4 \pi u_1$.

The mass outflow rate can also be evaluated at the inner boundary 
$\xi = 1$. The mass outflow rate must be the same in both limits
$\xi \to 1$ and $\xi \to \infty$, so this calculation provides a
consistency check. In this case, the dimensionless outflow rate is 
given by the expression
\be
{\dot m}_1 = \int_S \left( \alpha {\bf u} \right) \cdot \rhat dS \, , 
\ee
where the integral is taken over the entire planetary surface;
however, the outflow velocity ${\bf u}$ is only nonzero over the
fraction of the surface that supports the outflow. Notice also that
the flow is not radial, so that the velocity ${\bf u}$ points in the
$\phat$ direction (rather than the $\rhat$ direction). Using the
divergence theorem, the integral can be rewritten in the form 
\be
{\dot m}_1 = \int_V \nabla \cdot \left( \alpha {\bf u} \right) dV 
= \int_V {1 \over h_p h_q h_\phi} {\partial \over \partial p} 
\left( \alpha u h_q h_\phi \right) dV \, , 
\ee
where $V$ is the volume of the planet and where $u$ is the only
non-vanishing component of the velocity (in the $\phat$ direction).
Here, the volume element $dV = h_p h_q h_\phi dp dq d\phi$. 
The integral with respect to $\phi$ produces a factor of $2 \pi$ 
because the system is axisymmetric; the integral over $p$ can be 
evaluated directly and results in a surface term evaluated at the 
planetary surface $\xi$ = 1. The remaining expression becomes  
\be
{\dot m}_1 = 4 \pi \int_0^{q_X} u_1 \, q \, dq \, H_1^{-1/2} \, , 
\ee
which has the same form as that in the limit $\xi \to \infty$
(compare with equation [\ref{largemdotint}]). Again we must consider
flow from both the northern and southern hemispheres of the planet.

\begin{figure} 
\figurenum{5} 
{\centerline{\epsscale{0.90} \plotone{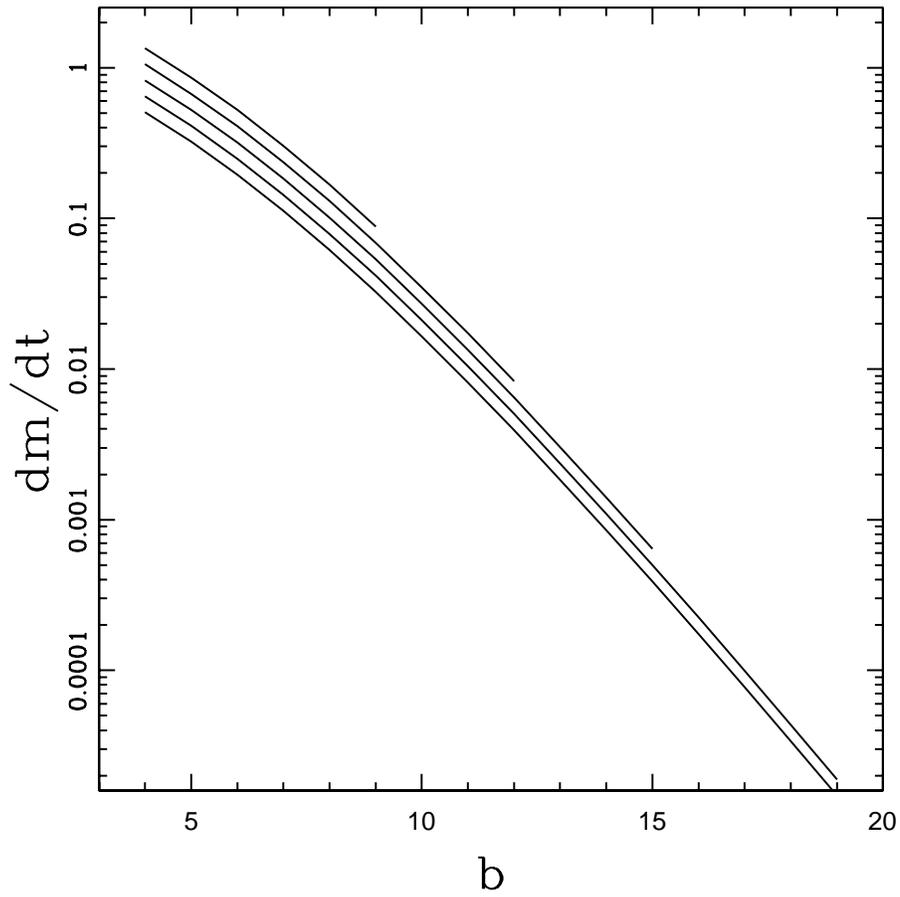} }}
\figcaption{Dimensionless mass outflow rate versus dimensionless 
depth $b$ of the gravitational potential well of the planet. The 
curves show the results for different values of the magnetic field 
strength parameter $\bstar$, from $\bstar$ = 0.0005 (bottom curve) to 
$\bstar$ = 0.008 (top curve). } 
\label{fig:mdot} 
\end{figure}
\begin{figure} 

\figurenum{6} 
{\centerline{\epsscale{0.90} \plotone{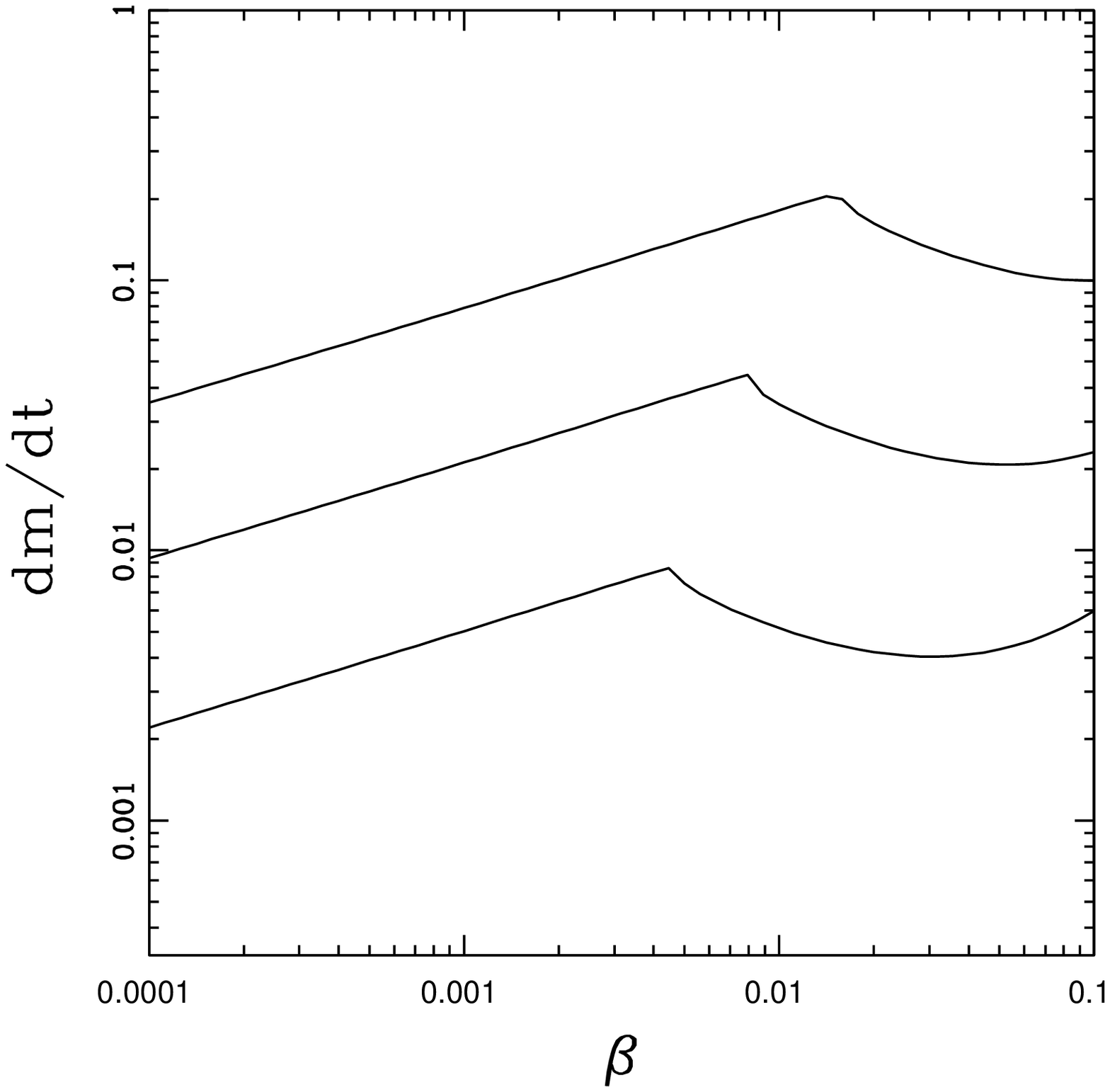} }}
\figcaption{Dimensionless mass outflow rate versus magnetic field 
strength parameter $\bstar$.  The curves show results for different
values of the dimensionless depth $b$ of the gravitational potential
well of the planet, where $b$ = 12 (bottom), $b$ = 10 (middle), and 
$b$ = 8 (top). Local maxima occur for the largest values of $\bstar$
that allow the flow along all of the open streamlines to pass smoothly 
through the sonic point (see text). } 
\label{fig:mdotalt} 
\end{figure}

Figure \ref{fig:mdot} shows the dimensionless mass outflow rate 
${\dot m} = dm/dt$ plotted as a function of the dimensionless depth
$b$ of the gravitational potential well of the planet. The results are
shown for a series of values of the field strength parameter $\bstar$;
the curves correspond to increasing values of $\bstar$ from bottom to
top in the figure. Notice that the curves end at particular values of
$b$, and that these values decrease with increasing $\bstar$.  For
larger values of $b$, only a fraction of the open streamlines allows
the flow to pass smoothly through the sonic point. The mass outflow
rates are thus diminished. In the discussion below we derive a scaling
law (see equation [\ref{mdotscale}]) that describes how the outflow
rate ${\dot m}$ depends on the variables $(b,\bstar)$; over the
parameter space represented in Figure \ref{fig:mdot}, this scaling
law holds to an accuracy of $\sim 3\%$.

Figure \ref{fig:mdotalt} provides another way to view the dependence
of the dimensionless outflow rates on the underlying parameters. In
this case, we plot the outflow rate ${\dot m}$ as a function of the
field strength ratio $\bstar$. Each curve corresponds to a different
value of the dimensionless depth $b$ of the gravitational potential
well of the planet. The outflow rates increase with $\bstar$ until a
well-defined maximum is reached; note that the parameters of this
extremum are defined by solutions to equation (\ref{failpoint}).  At
this point, further increases in $\bstar$ lead to straighter magnetic
field lines (and streamlines), especially along the poles of the
system, and the flow along those streamlines cannot pass smoothly
through the sonic point.  As $\bstar$ increases, the fraction of the
open streamlines that allow for smooth flow decreases, but the
fraction of streamlines that are open increases. In addition, the
sonic point moves inward, for those streamlines where it exists, and
this effect acts to increase the outflow rate ${\dot m}$.  These
competing effects thus lead to the non-monotonic behavior shown in
Figure \ref{fig:mdotalt}.

The parameter $\bstar$ is typically small; for example, 
$\bstar \sim 10^{-3}$ when the field strengths are equal on the stellar
and planetary surfaces and $a$ = 10 $R_\ast \approx 0.05$ AU.  As a
result, it is useful to find simplified results that are correct to
leading order in $\bstar$. If we expand equation (\ref{mdotexpress}),
the leading order term becomes
\be
{\dot m} = 3 \pi \langle u_1 \rangle \bstar^{1/3} + 
{\cal O} \left( \bstar^{2/3} \right) \, . 
\ee
Note that in the case of a spherically symmetric flow, the
dimensionless mass loss rate would have the form 
${\dot m}_{\rm sph} = 4 \pi u_1$ in these units. The fraction of 
the total possible outflow rate (with fixed $u_1$ = 
$\langle u_1 \rangle$) is thus $F \approx 3 \bstar^{1/3}/4$.  
From the previous section, the fiducial value of the parameter
$(\lambda/q)_0 \sim u_1 \sim b^3 \exp[-b]$ (see equation
[\ref{lambdafidu}]). We thus except the dimensionless mass outflow 
rate ${\dot m}$ to obey the scaling relation 
\be
{\dot m} \approx A_m \, b^3 \, \exp[-b] \, \bstar^{1/3} \, , 
\label{mdotscale} 
\ee
where $A_m$ is a constant of order unity. Fitting to the results
presented in Figures \ref{fig:mdot} and \ref{fig:mdotalt}, we
find $A_m \approx 4.8 \pm 0.13$ (where the quoted uncertainties
represent the standard deviation of $A_m$ for the parameter space
depicted in the Figures). This scaling law works well as long as the
flow along all of the open field lines can pass through the sonic
point. For sufficiently large $\bstar$, however, sonic transitions
cannot take place in the polar direction, and the dependence of the
outflow rate ${\dot m}$ on the parameters $(b,\bstar)$ becomes more
complicated. 

In the limit of large $\bstar$, sonic transitions cannot take place
along any of the directions. On the other hand, all of the field lines
from the planet must match onto stellar field lines. In this case, the
values of $\lambda/q$ are given by equation (\ref{lambdafail}) for all
of the streamlines. The integral that defines the dimensionless
outflow rate can be evaluated to obtain the result
\be
{\dot m} = 6 \pi \bstar^{-2/3} \exp \left[ - (b + 1/2) \right] \, . 
\label{mdotfail} 
\ee
This expression is only valid for $\bstar < 1$. At larger stellar
field strengths, all of the field lines originating on the planet
surface are open, and the leading coefficient in equation
(\ref{mdotfail}) becomes $2 \pi (1 + 2/\bstar)$. In the extreme limit
$\bstar \gg 1$, the dimensionless outflow rate thus reduces to the
expression ${\dot m} = 2 \pi \exp[-(b + 1/2)]$, which is the form
expected for flow in the $\zhat$ direction from a disk with the radius
of the planet.

\subsection{Estimating the Physical Constants} 

The previous subsections specify the solutions for the dimensionless
fluid fields, including the necessary conditions for passing smoothly
through the sonic point and specification of the dimensionless mass
outflow rate ${\dot m}$.  In this section, we complete the solution by
estimating values for the physical parameters $\rho_1$ and $a_s$ that
determine the full mass outflow rate, where ${\dot M}$ = $\rho_1 a_s
R_P^2 {\dot m}$.

We first provide order of magnitude estimates: The planetary radius is
given, with a typical value $R_P \approx 10^{10}$ cm. Since these
planets are gas giants, they do not have solid surfaces. As a result,
the planetary radii measured by transits are determined by the levels
in the atmosphere that are opaque to optical photons, typically at
pressures $\sim1$ millibar (see, e.g., Charbonneau et al. 2007).  We
also note that the launching radius where $\xi$ = 1 could lie several
scale heights above the planetary surface defined by transit
measurements (e.g., at pressures $\sim1$ nanobar). However, the scale
height $H \approx R_P/b$, where $b \sim 10$, or larger.  As a result,
$H \ll R_P$, and any departures of the launching radius from $R_P$ are
expected to be small ($\sim$10 percent).  The gas temperature is
expected to approach the benchmark value $T \sim 10^4$ K, so that
sound speed $a_s \sim 10$ km/s. Finally, the density $\rho_1$, or
equivalently $n_1$ = $\rho_1/m_P$, can be estimated by using the fact
that the wind is launched near the $\tau$ = 1 surface. The optical
depth $\tau \sim \sigmauv n_1 H$, so the number density $n_1 \sim 1 /
(H \sigmauv) \approx b/(R_P \sigmauv) \sim 10^9$ cm$^{-3}$.  With
these values, the scale for the mass outflow rate is ${\dot M}_{\rm
  scale} = m_P n_1 a_s R_P^2 \sim 10^{11}$ g/s (see also MCM). Since 
the dimensionless factor ${\dot m} \sim 0.01$ (see Figures
\ref{fig:mdot} and \ref{fig:mdotalt}, and equation [\ref{mdotscale}]),
typical outflow rates are expected to be ${\dot M} \sim 10^{9}$ g/s.

To obtain a better estimate for the quantities $(a_s, n_1)$, we need
to consider heating and cooling of the gas (equations [\ref{energy}],
[\ref{heating}], and [\ref{cooling}]) including ionization (equation
[\ref{ionbalance}]). Unlike previous outflow studies, the geometry of
the flow is determined by the magnetic field structure. For relatively
large $\bstar$, the magnetic field of the star dominates, and the
streamlines become primarily vertical. In the opposite limit of small
$\bstar$, only the streamlines from the polar regions lead to outflow,
and these streamlines are also oriented mostly in the $\zhat$
direction (see Figure \ref{fig:sonicsurf}). In either case, the flow
is directed in the polar directions, whereas stellar heating arrives
from the equatorial direction. Because of this configuration, the
optical depth of the incoming radiation is not directly tied to the
mass outflow rate (which occurs for spherical flow). In addition, the
stellar UV photons penetrate (from the side) into the atmospheric
layers where the flow speeds are small. Leaving a full
three-dimensional treatment of the heating/cooling problem for the
future, we adopt here a simplified approach where flow velocities are
neglected. As a further approximation, we assume that the required
values of the sound speed $a_s$ and density $n_1$ are determined at
the layer where the optical depth is unity, where $n_1$ = $\rho_1/m_P$
is the total number density at the $\tau$ = 1 surface.

In this limit, the energy equation (\ref{energy}) reduces to the
condition that heating and cooling are locally in balance, so that
$\Lambda = \Gamma$. Similarly, the ionization equation
(\ref{ionbalance}) reduces to the statement that the rate of
ionization balances the recombination rate. These two equations can be
combined to determine the ionization fraction as a function of
temperature. With the definition $X_{+} \equiv n_{+} / n_0$, we can
write
\be
X_{+} (T) = { n_{+} \over n_0} = 
{C \over \effabsorb \langle h \nu \rangle \alpha_R} 
\exp[-T_C/T] \approx \, (2.5 \times 10^5) \, \, T_4^{0.9} \, 
\exp[-11.8348/T_4] \, . 
\ee
With the ionization fraction specified, the heating equation 
determines the temperature and can be written in the form  
\be
\exp[-T_C/T] {X_{+} \over 1 + X_{+}} = 
{\effabsorb F_{UV} \sigmauv \over C n} \exp[-\tau] =
{\effabsorb F_{UV} \sigmauv \over C {\rm e}} n_1^{-1} \, , 
\label{temp} 
\ee
where $n = n_0 + n_{+}$ is the total number density and $n_1$ =
$n(\tau=1)$. The left hand side of equation (\ref{temp}) is a function
of temperature, whereas the right hand side is a function of position
only. The second equality specializes to the layer where $\tau$ = 1. 
If we assume that the flow velocities are small, the optical depth
is given by
\be
\tau = {n_1 \sigmauv R_p \over b (1 + X_{+})} = 1 \, , 
\label{tau} 
\ee
where $b$ is the dimensionless depth of the gravitational potential
well (and note that $b \propto 1/T$).  Equations (\ref{temp}) and
(\ref{tau}) provide two equations for the two unknowns $n_1$ and 
$T$. With these parameters specified, we can then evaluate the scale 
${\dot M}_{\rm scale} = m_P n_1 a_s R_P^2$ for the mass outflow rate. 
The outflow rate itself is given by ${\dot M}= {\dot m} (m_P n_1 a_s
R_P^2)$, where the dimensionless outflow rate ${\dot m}$ can be
approximated using the scaling law from equation (\ref{mdotscale}) or
the full calculation in that section. Note that in order to determine
the value of $b$ = $G M_P / a_s^2 R_P$, we need the temperature 
$T \propto a_s^2$. 

\begin{figure} 
\figurenum{7} 
{\centerline{\epsscale{0.90} \plotone{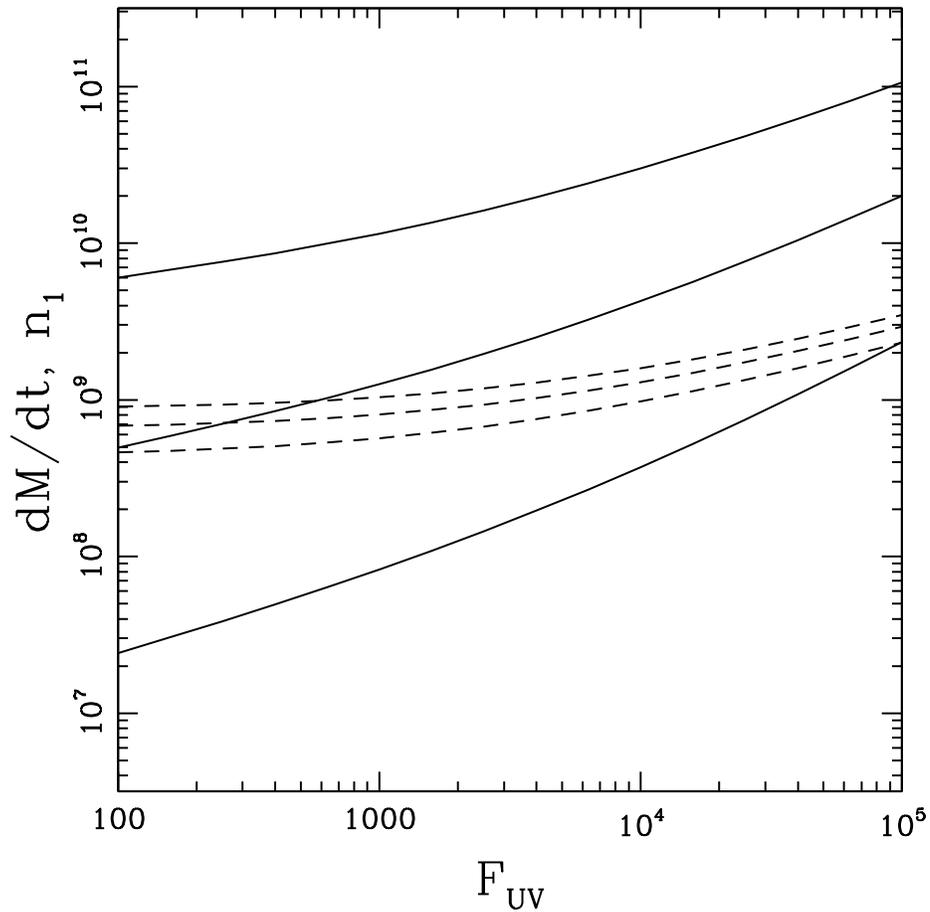} }}
\figcaption{Mass outflow rates as a function of UV flux from the 
star. The outflow rate $dM/dt$ = ${\dot M}$ is given in g/s, and the
UV flux is given in erg cm$^{-2}$ s$^{-1}$. Solid curves show the
outflow rates for three values of the planet mass, from $M_P$ = 0.5
$M_J$ (top curve) to $M_P$ = 1.0 $M_J$ (bottom curve). Dashed curves
show the number density $n_1$ (in cm$^{-3}$) at the $\tau$ = 1 surface
for the same cases. The magnetic field strength parameter $\bstar$ =
0.001 for all cases shown. }
\label{fig:mdotphys} 
\end{figure}

Figure \ref{fig:mdotphys} shows the number densities $n_1$ and
corresponding outflow rates ${\dot M}$ as a function of the UV flux
$F_{UV}$ for three choices of planetary mass: $M_P$ = 0.5, 0.75, and
1.0 $M_J$. Keep in mind that the dimensionless outflow rate ${\dot m}$
depends on the magnetic field strength parameter $\bstar$ (see Figure
\ref{fig:mdotalt}); these curves are calculated using $\bstar$ = 0.001.
In this approximation, the outflow rate increases somewhat more slowly
with $F_{UV}$ than the linear relation of equation
(\ref{mdotestimate}). For sufficiently large flux levels (not shown),
the temperature approaches its effective maximum value (just above
$10^4$ K, see Spitzer 1978); in this regime, the outflow rates would
approach a constant value, but the assumptions of this section break
down.  Notice also that the number density (dashed curves) is a slowly
varying function of both the UV flux and the planet mass, and that
$n_1 \sim 10^9$ cm$^{-3}$ as expected.

\section{Observational Signatures} 
\label{sec:observe} 

Observations used to infer the presence of planetary outflows show
that the transit depth is greater at UV wavelengths than in the
optical (Vidal-Majar et al. 2003, 2004; D{\'e}sert et al. 2008, Sing
et al. 2008, Lecavelier des Etangs et al. 2008, Linsky et al. 2010).
As a result, we need to determine the optical depth of the outflow to
UV radiation. To start, we define the dimensionless column density 
$N_{\rm c}$ according to
\be
N_{\rm c} \equiv \int_{-\infty}^\infty \alpha \, ds \, \, , 
\label{column} 
\ee
where $\alpha$ is the dimensionless density obtained from the flow
solution (see Section \ref{sec:launch}). The column density 
$N_{\rm c}$ is thus defined for a given path and the variable $s$ 
is the dimensionless distance along the path. 

Note that equation (\ref{column}) determines the total column density,
whereas the optical depth depends on the chemical species that absorbs
the UV radiation. Most observations to date are carried out for
Lyman-$\alpha$ photons, so that the column density of interest is that
of neutral hydrogen. As a result, ionization modeling is necessary to
determine the optical depths and hence the observational signatures.
This present treatment assumes isothermal flow. To consistent order of
approximation, the neutral column density is reduced from that of
equation (\ref{column}) by a factor ${\cal F} = (1 + X_{+})^{-1}$ (see
equations [\ref{temp}] and [\ref{tau}]). For the benchmark temperature
$T = 10^4$ K, $X_{+} \approx 1.8$, so that the neutral hydrogen column
density is reduced to a fraction ${\cal F} \approx 0.36$ of the total
column density. However, the ionization fraction is a sensitive
function of temperature.  For $T = 8000$ K, $X_{+} \approx 0.08$, and
the factor ${\cal F} \approx 0.93$.
 
\begin{figure} 
\figurenum{8} 
{\centerline{\epsscale{0.90} \plotone{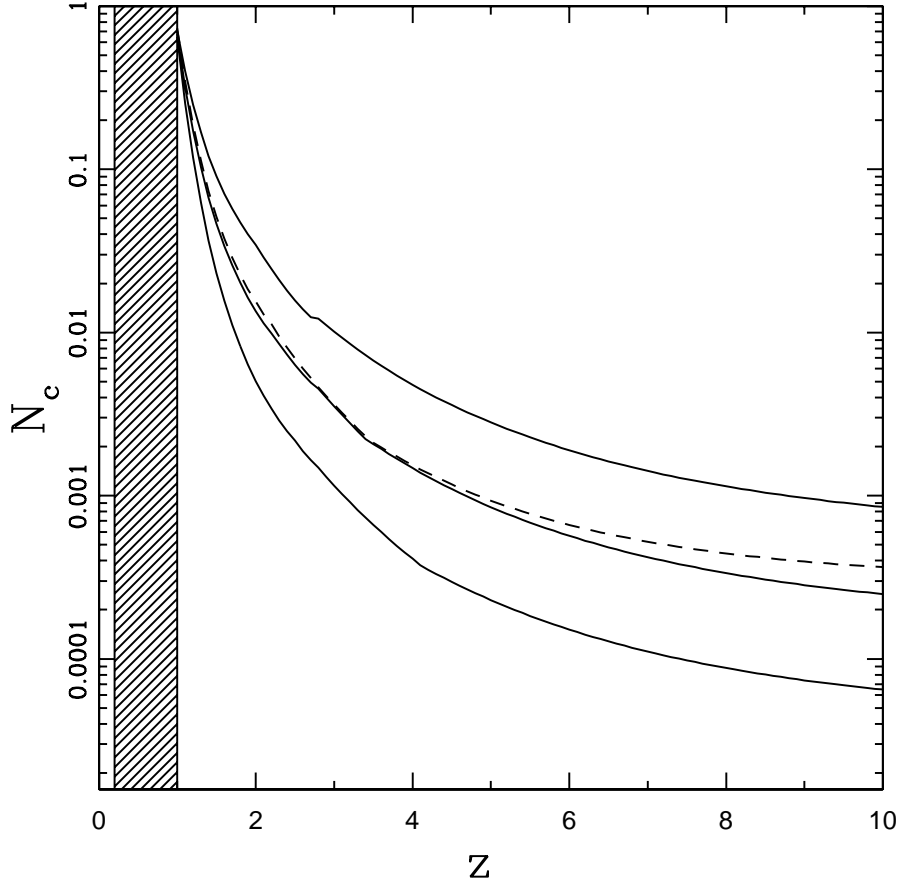} }}
\figcaption{Dimensionless column density through the outflow as a 
function of vertical coordinate $z$ (in units of the planet radius 
$R_P$).  Solid curves show the column density for magnetic field
parameter $\bstar$ = 0.001 with three values of the dimensionless depth
$b$ of the planetary gravitational potential well: $b$ = 8 (top), 10
(middle), and 12 (bottom). The dashed curve shows the $b$ = 10 case
with larger field strength parameter $\bstar$ = 0.002. Optical depth
is obtained from the dimensionless column density through the relation 
$\tau = {\cal F} n_1 \sigmauv R_P N_{\rm c} \sim 20 N_{\rm c}$. } 
\label{fig:column} 
\end{figure}

For the sake of definiteness, we evaluate the column density on paths
that are parallel to the star-planet direction ($y$ = 0 in the
coordinate system used here; see Figure \ref{fig:blines}) with a
constant vertical height $z$. Each point along the path corresponds to
a different streamline, and hence a different value of the constant
$\lambda/q$.  Figure \ref{fig:column} shows the dimensionless column
density $N_{\rm c}$ through the outflow, as a function of $z$, for
several typical cases.  The solid curves show the effect of varying
the depth of the gravitational potential well of the planet.  The
three curves correspond to $b$ = 8, 10, and 12, where the magnetic
field strength parameter $\bstar$ = 0.001, which corresponds to equal
surface fields on the star and planet (for $a$ = 10 $R_\ast$).  The
dashed curve shows the effect of doubling the field strength parameter
$\bstar$ (for $b$ = 10).

The optical depth at UV wavelengths is given by $\tau = {\cal F} n_1
\sigmauv R_P N_{\rm c}$. For typical cases, the quantity $n_1 \sigmauv
R_P \sim 20$. The base density $n_1 = \rho_1/m_P$ is a slowly
increasing function of the incident UV flux $F_{UV}$ (see Figure
\ref{fig:mdotphys}), but the correction factor ${\cal F}$ due to
ionization decreases with temperature and hence decreases with
increasing $F_{UV}$.  The range in $z$ for which the flow is optically
thick is thus relatively small, ranging from $z/R_P$ = 1.35 to 1.77
for the cases shown with $b$ = 8 to 12. With these values, the area
for which the planet is apparently optically thick will be 2 -- 3
times larger at UV wavelengths than in the optical bands. The transits
for UV observations are thus predicted to be 2 -- 3 times deeper.

\begin{figure} 
\figurenum{9} 
{\centerline{\epsscale{0.90} \plotone{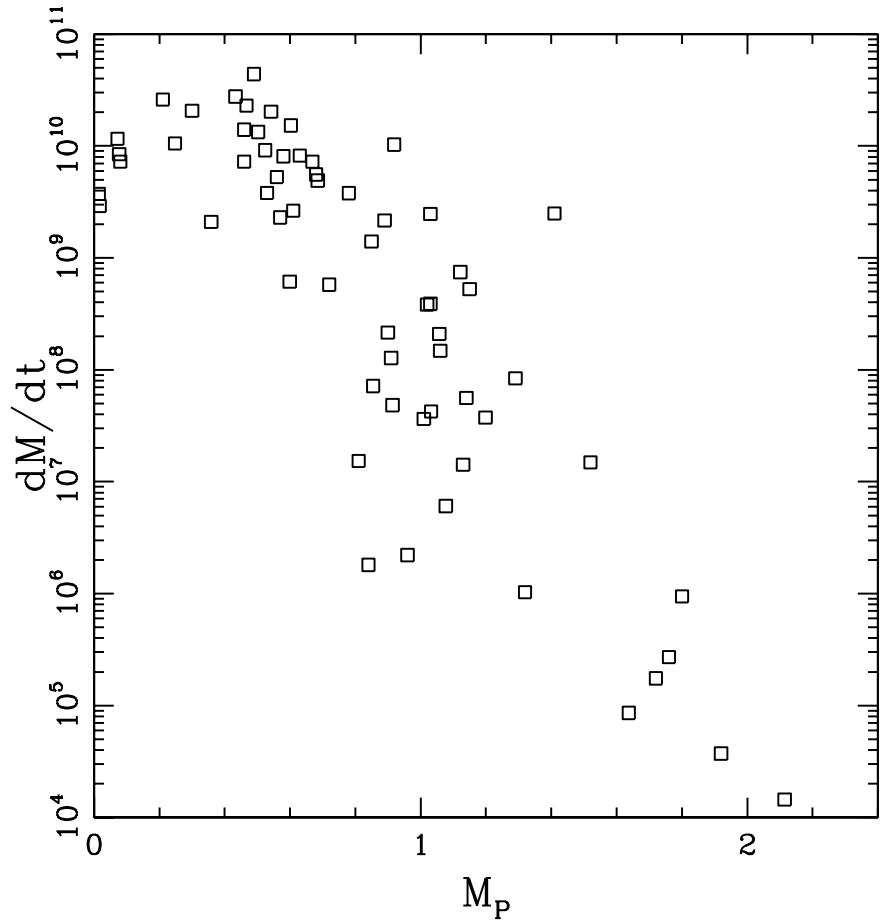} }} 
\figcaption{Rough estimates for mass outflow rates for the sample 
of extrasolar planets observed in transit. Mass outflow rates $dM/dt$
= ${\dot M}$ are given in g/s; planetary masses are given in Jovian
masses $M_J$. }
\label{fig:observemdot} 
\end{figure}

The mass outflow rates are expected to vary substantially from planet
to planet. In order to illustrate this trend, in Figure
\ref{fig:observemdot} we plot estimates for the mass outflow rates for
the collection of extrasolar planets that are observed in transit
(data from Schneider 2010). The planetary masses, radii, and semimajor
axes are observed.  In order to estimate the outflow rates, we must
also specify the UV flux $F_{UV}$ and the magnetic field strength
parameter $\bstar$. With these quantities determined, the results of
Section \ref{sec:launch} can be used to calculate the mass outflow
rate ${\dot M}$. The UV fluxes from main-sequence stars vary with
spectral type and other stellar parameters (e.g., rotation rates) in a
complicated manner (see Lecavelier des Etangs 2007, Lammer et al.
2003, and references therein). To obtain the results shown in Figure
\ref{fig:observemdot}, we estimate the UV flux using a simple scaling
law that is intermediate between the scaling laws advocated by the
aforementioned authors. To specify the magnetic flux parameter
$\bstar$, we assume that the stellar and planetary surface fields are
comparable, so that to leading order $\bstar \propto a^{-3}$. Finally,
we take into account the fact that radius of the $\tau$ = 1 surface
(where the outflow is launched and the parameter $b$ = $GM_P/a_s^2 R$
is evaluated) lies above the planetary surface at $R_P$. Using a
simple hydrostatic model for the lower layers of the outflow region,
we estimate that $R_1 \approx 1.2 R_P$, and this correction is used here.

With the specifications described above, we obtain the mass outflow
rates shown in Figure \ref{fig:observemdot} as a function of planet
mass. Several trends are clear: First, for a given planet mass, a wide
range of outflow rates are possible. Both the magnetic field parameter
$\bstar$ and the UV flux $F_{UV}$ vary with semimajor axis $a$, so
that closer planets are predicted to have stronger outflow rates. In
addition to this dependence on $a$, however, both variables ($\bstar$,
$F_{UV}$) are expected to display substantial variation from system to
system.  Next we note that range of outflow rates shown in Figure
\ref{fig:observemdot} spans seven orders of magnitude. Those planets
with the largest outflow rates should thus have observable transit
signatures, whereas those on the other end of the range will show
little or no effect.  Finally, the outflow rates are predicted to be a
steeply decreasing function of planetary mass $M_P$. In the isothermal
limit, we obtain the nearly exponential decrease shown in Figure
\ref{fig:observemdot} (see equation [\ref{mdotscale}]). If the gas can
cool substantially before it passes through the sonic transition, this
decrease could be less steep, but the overall trend remains.

Given that two observed planets have signatures of planetary outflows,
it useful to see how they compare with the results shown in Figure
\ref{fig:observemdot}. The planet HD209458b (Vidal-Majar et al.2003)
has mass $m_P$ = 0.685 $m_J$, and the figure shows that the expected
outflow rates are ${\dot M} \sim 10^{10}$ g/s. Even with the reduction
due to magnetic effects, such estimated outflow rates are large enough
to account for the observations of HD209458b (although these values
are near the low end of the inferred range).  However, for the planet
HD189733b (Lecavelier des Etangs et al. 2010), the mass is larger,
$m_P$ = 1.15 $m_J$, and the predicted outflow rate from this theory is
much lower. The values shown in Figure \ref{fig:observemdot}
correspond to ${\dot M} \sim 10^8$ g/s with significant scatter. The
observational papers for HD189733b advocate a large UV flux (up to 40
times the solar value), which increases the theoretical outflow rate
to ${\dot M} \sim 10^9$ g/s for our choice of field strength parameter
$\bstar = 10^{-3}$. To account for the inferred outflow rate of 
${\dot M} \sim 10^{10}$ g/s, one could invoke a larger value of
$\bstar$ for this system. Nonetheless, some tension remains between 
the inferred outflow rate and that expected theoretically from 
magnetically controlled models. More observations, of this system 
and others with a range of masses, are need to sort out this
comparison.

\section{Conclusion}
\label{sec:conclude} 

\subsection{Summary of Results} 

This paper has begun a theoretical study of outflows from the surfaces
of Hot Jupiters in the regime where the flow is controlled by magnetic
fields. In this case, the magnetic field structure determines the flow
geometry (whereas the field configurations are determined by
independent dynamo processes within the planet, and by the background
contribution from the star). With the magnetic field structure
specified, the dimensionless version of the outflow problem can be
solved semi-analytically; this paper carries out this calculation in
the isothermal limit, including the requirement that the flow must
pass smoothly through the sonic transition.  The determination of the
physical constants represents the final piece of the calculation. The
specific results of this paper can be summarized as follows:

[$\ast$] This paper considers a reduced description of the magnetic
field structure that includes the dipole field of the planet and a
constant background contribution.  The reduced field structure near
the planet is modeled in Section \ref{sec:field}, and provides a
description of the flow geometry in the region where the outflow is
launched (see Figure \ref{fig:sonicsurf}). In this section we
construct the corresponding coordinate system and differential
operators for this flow geometry.  These results are used here to
study the launch of outflows for Hot Jupiters, but can also be used in
a variety of other applications.

[$\ast$] A fraction $F_{AP}$ of the planetary surface supports
magnetic field lines that are open with respect to the planet. This
area fraction provides a constraint on the mass flow that can fully
leave the planet. The remaining fraction, 1 -- $F_{AP}$, defines the
region where material can leave the planetary surface but is
nonetheless confined to the immediate vicinity of the planet; this
material produces an exosphere surrounding the planet.  In the
magnetically controlled regime, the outflow rate is thus lower than
estimates obtained by assuming spherical symmetry. The fraction
$F_{AP}$ is calculated analytically for the reduced field
configuration near the planet (Section \ref{sec:field}).

[$\ast$] Along each streamline, the dimensionless energy $\varepsilon$
and the flow momentum parameter $\lambda$ are constant, but vary
across streamlines. These parameters are specified by requiring that
the flow pass smoothly through the sonic point, and by the boundary
conditions at the planetary surface. The resulting specification of
parameters can be found analytically (see equations [\ref{matchingtwo}] 
and [\ref{lambda}], and Figure \ref{fig:lambda}). The outflow rate is
determined by integrating over the surface area of the planet where
the outflow is active, where the angle of the flow direction (which 
is not radial) must be included (see equations [\ref{mdotfirst} -- 
\ref{mdotscale}]). The resulting mass outflow rates are well-defined
functions of the depth $b$ of the planetary potential well and the
magnetic field strength parameter $\bstar$ (see Figures \ref{fig:mdot}
and \ref{fig:mdotalt}). Over the regime of parameter space where flow
along all of the open field lines can pass smoothly through the sonic
point, the scaling law of equation (\ref{mdotscale}) provides an
accurate determination of the dimensionless outflow rate ${\dot m}$.

[$\ast$] This flow geometry is significantly different from previous
cases that assume spherical symmetry. For magnetically controlled
outflows, only a fraction of the field lines (and hence streamlines)
allow outflow. In addition, passage through the sonic point depends
sensitively on the divergence operator (at the transition) which
depends on the configuration of the streamlines. As the magnetic field
strength of the background (from the star) increases relative to that
of the planet, more of the surface has open field lines, but not all
of the streamlines allow for smooth sonic transitions.  Another key
difference is that spherical flows are often taken to be
``self-limiting'', where the outflow rate, in part, determines the
optical depth to the incoming photons. In this geometry, most of the
heating photons impinge upon the system from the equatorial
directions, whereas most of the outflow is directed along the poles of
the planet. This flow geometry also changes the effect of tidal forces
exerted on the planet by the star: Because mass loss occurs along the
polar directions, the tidal forces act to inhibit, rather than
enhance, the mass outflow rates (see also TAL; compare with MCM).

\subsection{Discussion} 
\label{sec:condiscuss} 

Since this problem contains a number of physical parameters, it is
useful to summarize them here and discuss which ones are the most
important. The structure of the magnetic field requires the
specification of three quantities: the surface field strength $B_P$,
the effective planetary radius $R_P$, and the ratio $\bstar$ of the
background field strength to the surface field strength. For the cases
under consideration, however, the magnetic field is assumed to be
strong enough to guide the flow, so that only the field geometry plays
a role. In this simplified treatment, the magnetic field geometry is
characterized by the single parameter $\bstar$ (see equation
[\ref{reducedfield}]). For purposes of launching the wind from the
planet, the main contribution of the stellar field is to provide a
nearly constant background field, taken here to lie in the vertical
direction.  Since the planetary field is much stronger in the region
within the sonic surface, this assumption is valid for determining the
launch of the wind. However, the stellar field direction will not
necessarily line up with the pole of the planet, so that more general
geometries should be considered in the future.  In addition, the
stellar field configuration will affect the manner in which the
outflow propagates after its launch --- beyond the sonic surface ---
and this problem should also be addressed in future work.

With the flow geometry set by the magnetic field structure, the
outflow problem determines the fluid variables as a function of the
(single) coordinate $p$ which follows the field lines (Section
\ref{sec:launch}). These quantities include the density $\rho(p)$,
flow speed $v(p)$, temperature $T(p)$, and ionization fraction 
$X_{+} (p) = n_{+}/n_0$, and depend on the planet mass $M_P$ and
radius $R_P$, and the stellar heating flux $F_{UV}$. Under the
assumption of isothermal flow, the problem is reduced further to 
two dimensionless fluid variables $\alpha$ = $\rho/\rho_1$ and 
$u = v/a_s$, along with the dimensionless depth of the planetary
gravitational well $b$ = $GM_P/(a_s^2 R_P)$. The dimensionless problem
thus has only two parameters $(b,\bstar)$, and they determine the
dimensionless outflow rate (see equations [\ref{mdotfirst}] --
[\ref{mdotfail}] and Figures \ref{fig:mdot} and \ref{fig:mdotalt}).  

The determination of physical quantities requires specification of the
density scale $\rho_1$, the sound speed $a_s$, and the planet radius
$R_P$. The UV flux $F_{UV}$ from the star determines, in part, the
sound speed $a_s$ and the density $\rho_1$ at the base of the flow.
Over the expected parameter space, the sound speed $a_s$ and density
scale $n_1$ vary by only factors of 3 -- 10. However, the outflow rate
decreases exponentially with the depth $b$ of the potential well, and
shows complicated dependence on the magnetic field strength parameter
$\bstar$ (see Figures \ref{fig:mdot}, \ref{fig:mdotalt}, and
\ref{fig:mdotphys}). Note that this exponential sensitivity to the
depth of the potential well is a common feature in outflow problems
where the sound speed is less than the escape speed (compare with the
case of outflows from circumstellar disks driven by external FUV
radiation; Adams et al. 2004). Since the dimensionless outflow rate
displays exponential dependence on $b$, where $b$ = $G M_P / R a_s^2$,
the exact value of the temperature (or equivalently the sound speed)
can be important. In addition, the radius $R$ in this expression
corresponds to the radius $R_1$ where the outflow is launched, i.e.,
the $\tau = 1$ surface. Although the the radius $R_1$ differs from the
planet radius $R_P$ by only $\sim10$ percent, this difference can be
significant for the regime of large $b$ where the outflow rates are
exponentially suppressed.

\subsection{Future Work} 

This paper represents only the first step toward understanding
outflows from Hot Jupiters in the regime where magnetic fields
dominate the flow geometry. This work should be carried forward 
in a number of directions: 

One approximation used here is the assumption of isothermal flow. For
the next stage of development, an analogous calculation can be carried
out using a more general, polytropic equation of state.  However, a
full treatment of the heating and cooling should be undertaken. This
calculation requires one to solve the energy equation (\ref{energy})
and the ionization equation (\ref{ionbalance}), where the heating from
the central star is determined through a three-dimensional radiative
transfer calculation.  The chemistry of the outflow should also be
included, both to get a better description of the heating and cooling
processes and to determine observational signatures (e.g., Garc{\'i}a
Mu{\~n}oz 2007). Note that one of the intrinsic complications that
arises in this problem is that the flow configuration (determined by
magnetic field lines) does not have the same geometry as the heating
and cooling processes. 

This paper considers a dipole magnetic field on the planet and a
constant background field from the star; in addition, this work
specializes to the case where the stellar contribution to the magnetic
field near the planet is purely vertical. A wide range of magnetic
field configurations are possible in star/planet systems and these
possibilities should be explored further. One particular issue arises
with sufficiently strong stellar fields, which provide a background
field for purposes of launching the wind from the planet. If the
stellar field is sufficiently strong and straight, the flow cannot
pass throughly through a sonic transition.  In this case, the sonic
point is effectively removed to spatial infinity and the outflow rates
are suppressed (e.g., see equation [\ref{mdotfail}]). For the geometry
considered here this suppression arises when the field strength
parameter $\bstar \gta 0.01$ (which requires the stellar surface field
to be $\sim10$ times that of the planet for $a$ = 0.05 AU, or equal to
that of the planet for $a$ = 0.023 AU).  A more detailed study of this
regime should be undertaken, including more complicated configurations
for the background (stellar) field. Another related effect is that the
interaction between the planetary magnetic field and the stellar
magnetosphere, including the flow of material considered here, can
lead to an enhancement in stellar activity (e.g., Cuntz et al. 2000,
Cohen et al. 2009) and orbital evolution (Chang et al. 2010).  This
paper also focuses on the case where the stellar field is vertical,
which would arise, e.g., from a dipole field from the star that
rotates with the same angular velocity as the planetary orbit. Future
work should relax this assumption. In particular, the field lines
could trail the planet, wrap up, and lead to further complications.

For sufficiently strong magnetic fields and low outflow rates, the ram
pressure of the flow is not strong enough to affect the underlying
field structure. For weaker fields and/or higher outflow rates,
however, the back-reaction of the flow on the magnetic field should be
taken into account. An understanding of this physics can be attained
through the solution to the Grad-Shafranov equation (e.g., Shafranov
1966), which self-consistently determines the distribution of
streamlines (see also Cai et al. 2008). Even for the strong field
limit, considered here, this back-reaction plays a role near the
X-point. On a related note, the closed field lines can also be loaded
with mass and the planet will thus develop a quasi-static exosphere
(see TAL); this exosphere can enhance the observational signature, 
by increasing the transit depth at UV wavelengths, and should thus 
be considered further.

Finally, we note that some of the planets observed in transit have
orbits with nonzero eccentricity.  The canonical example is the planet
HD17156b (Barbieri et al. 2007), which has an orbital eccentricity of
$e \approx 0.67$; in this system, the stellar flux varies by a factor
of $\sim25$ over the course of the planetary orbit.  The planetary
outflow in this system, and others with similar architecture, will
thus be time dependent. The resulting time dependent outflow rates
will be sensitive functions of the heating and cooling mechanisms, as
well as the (complicated and time dependent) magnetic field
configurations. The study of such systems will provide sensitive
tests of the outflow mechanism.

$\,$

\acknowledgments 

This paper benefited from discussions with many colleagues, especially
Phil Arras, Mike Cai, Daniele Galli, Zhi-Yun Li, Greg Laughlin, Susana
Lizano, Nathan Schwadron, and Frank Shu. An anonymous referee provided
many useful comments and suggestions. This work was supported at the
University of Michigan through the Michigan Center for Theoretical
Physics. Portions of this work were carried out at the Kavli Institute
for Theoretical Physics at the University of California, Santa Barbara
(supported by the National Science Foundation under Grant No.
PHY05-51164).  FCA is supported by NASA through the Origins of Solar
Systems Program (grant NNX07AP17G) and by NSF through the Division of
Applied Mathematics (grant DMS-0806756).

\newpage 

\appendix 
\section{Equivalent Spherical Problem} 
\label{sec:equisphere} 

One complication inherent in the study of magnetically controlled
outflows is that the problem is three dimensional. Even the reduced
problem considered here, with a constant stellar field, requires two
dimensions. In order to gain further insight into the problem, this
Appendix develops an equivalent spherically symmetric version of the
wind problem.

The geometry of the flow pattern is specified by the divergence
operator. The quantity that appears in the conditions that match
solutions at the sonic point can be written as
\be
G_{\rm op} \equiv {\cal G} 
\left( {\partial \xi \over \partial p} \right)^{-1} \, , 
\ee 
where the geometrical factor ${\cal G}$ is given by equation 
(\ref{gdef}), and $\partial \xi/\partial p$ is given by equation 
(\ref{partialderiv}). In the limit of large $\xi \gg 1$, one can 
show that $G_{\rm op} \sim 12/(\bstar \xi^4)$. In the opposite limit 
where $\xi \sim 1$, near the planetary surface, $G_{\rm op} \sim 3/\xi$. 
This latter form is consistent with the result for a dipole field 
configuration near the poles of the system (recall that the outflow 
is concentrated near the poles). These two limiting forms can be 
connected through intermediate values by adopting the form 
\be
G_{\rm op} = {3 \over \xi (1 + \bstar \xi^3 / 4)} \, . 
\ee
The integrated form of the dimensionless continuity equation then becomes 
\be
\alpha u {\xi^3 \over 1 + \bstar \xi^3 / 4} = \lambda \, , 
\label{sphericalcont} 
\ee
where $\lambda$ is a (single) constant for the equivalent spherical
problem. The differential form of the continuity equation is obtained
by taking the derivative of equation (\ref{sphericalcont}). Near the
planet, we thus obtain $d(\alpha u \xi^3)/d\xi$ = 0, the form
appropriate for a dipole divergence. Far from the planet, the
continuity equation reduces to the form $d(\alpha u)/d\xi$ = 0, which
is the form applicable to flow along a single (cartesian) direction.
The force equation remains the same (see equation [\ref{simpleforce}]). 

With this form for the divergence, the condition for flow passing
smoothly through the sonic point takes the form
\be
b = {3 \xi \over 1 + \bstar \xi^3 / 4} \, , 
\label{sphericalmatch} 
\ee
which thus specifies the radius $\xi_s$ of the sonic transition
(compare with equation [\ref{failpoint}]). In the limit $\bstar \to 0$,
this matching condition becomes $b = 3 \xi$, as expected for a dipole.
This expression (\ref{sphericalmatch}) has no real solutions for
sufficiently large $b$ and/or large $\bstar$. The condition required
for solutions to exist, and hence for the flow to pass through the
sonic point, can be written in the form 
\be
\bstar b^3 < 16 \, . 
\ee
For cases where sonic transitions are possible, the required value 
of the constant $\lambda$ is thus given implicitly by the relation 
\be
{1 \over 2} \lambda^2 (1 + \bstar/4)^2 - \ln \lambda = {1 \over 2} + 
\ln (\bstar/4 + \xi_s^{-3}) + b (1 - 1/\xi_s) \, , 
\ee
where $\xi_s$ is given by the solution to cubic equation
(\ref{sphericalmatch}).  For consistency, the area subtended by the
outflow must be proportional to the function $\xi^3/(1 + \bstar
\xi^3/4)$ that appears in the continuity equation
(\ref{sphericalcont}). This quantity starts near unity at the planet
surface, grows like $\xi^3$ near the planet, and then approaches a
constant value in the limit $\xi \to \infty$. This behavior is thus
analogous to that of the more physical problem considered in the text.

\end{document}